\documentclass[preprint, 3p,12pt,authoryear]{elsarticle}

\usepackage{amssymb}
\usepackage{bm,subfig}
 \usepackage{amsthm,amsmath,mathrsfs}
\newtheorem{lem}{Lemma}
\renewcommand{\l}{\left(}
\renewcommand{\r}{\right)}
\usepackage{multirow}
\DeclareMathOperator*{\argmin}{argmin}
\usepackage{float}
\usepackage{comment}
\usepackage{natbib}

\begin{document}

\begin{frontmatter}

\title{Optimal classification and generalized prevalence estimates for diagnostic settings with more than two classes}

 \author[label1,label2]{Rayanne A. Luke\corref{cor1}} \ead{rluke3@jhu.edu}
 
 \author[label2]{Anthony J. Kearsley}
 \author[label2]{Paul N. Patrone}
 
\cortext[cor1]{Corresponding author} 

\address[label1]{Johns Hopkins University, Department of Applied Mathematics and Statistics, Baltimore, MD, 21218, USA}

\address[label2]{National Institute of Standards and Technology, Information Technology Laboratory, Gaithersburg, MD, 20899, USA}
 
% \affiliation[label1]{organization={Johns Hopkins University, Department of Applied Mathematics and Statistics},
        % city={Baltimore},
        % postcode={21218},
        %     state={MD},
        %     country={USA}}
%%
% \affiliation[label2]{organization={National Institute of Standards and Technology, Information Technology Laboratory},
          %  city={Gaithersburg},
          %   postcode={20899},
          %  state={MD},
          %   country={USA}}

\begin{abstract}

An accurate multiclass classification strategy is crucial to interpreting antibody tests. 
However, traditional methods based on confidence intervals or receiver operating characteristics lack clear extensions to settings with more than two classes.
 We address this problem by  developing a multiclass classification based on probabilistic modeling and optimal decision theory that minimizes the convex combination of false classification rates.  
 The classification process is challenging when the relative fraction of the population in each class, or  generalized prevalence, is unknown.  Thus, we also develop a method for estimating  the generalized prevalence of test data that is independent of classification. 
We validate our approach on serological data with severe acute respiratory syndrome coronavirus 2 (SARS-CoV-2)  na{\"i}ve, previously infected, and vaccinated classes. 
Synthetic data are used to demonstrate that (i) prevalence estimates are unbiased and converge to true values  
and (ii) our procedure applies to arbitrary measurement dimensions.
In contrast to the binary problem, the multiclass setting offers wide-reaching utility as the most general framework and provides new insight into prevalence estimation best practices.

\end{abstract}


\end{frontmatter}

\section{Introduction}

The severe acute respiratory syndrome coronavirus 2 (SARS-CoV-2) pandemic has highlighted the importance of accurate classification of antibody test results. Most work has focused on labeling data as previously infected (seropositive) or na{\"i}ve. 
Due to the deployment of SARS-CoV-2 vaccines in late 2020, there is a clear need for a classification scheme that correctly distinguishes between na{\"ive}, previously infected, and uninfected but vaccinated individuals. However, the traditional diagnostic classification methods of confidence intervals and receiver operating characteristics have no obvious extensions to a multiclass setting. 

 Current multiclass applications in 
  diagnostic classification are mostly limited to supervised learning and do not address the central role of mathematical modeling in
diagnostics.  Example studies include the application of support-vector machines to automatically sort endomysial autoantibody tests of celiac disease into one of four classes \citep{caetano2019automatic}, and another that trained deep neural networks to label resting-state functional magnetic resonance imaging results with one of six Alzheimer's disease state \citep{ramzan2020deep}.  
However, these approaches may not accurately quantify precise training population characteristics or account for the role of prevalence,  both of which should inform the classification procedure. In contrast, modeling can overcome this limitation. Binary (two class) examples include two-dimensional (2D) modeling of antigen targets coupled with optimal decision theory \citep{patrone2021classification}, statistical modeling applied to either antibody or viral-load tests   \citep{bottcher2022statistical}, and an approach to the time-dependent problem for antibody measurements \citep{bedekar2022prevalence}. However, none of these works discussed multiclass extensions. 

This paper uses mathematical modeling to fully address the task of multiclass classification in the context of diagnostic testing. We begin by showing that the notion of \textit{generalized prevalence}--the relative fraction of the population in each class--is fundamental for defining our objective function, the convex combination of false classification rates (Section \ref{sec:classif}). Minimization thereof yields optimal classification. Interestingly, we show that these prevalences can be computed without classification by solving a linear system. 
 We validate our methods using a SARS-CoV-2 serological data set with na{\"i}ve, previously infected, and vaccinated classes \citep{ainsworth2020performance,wei2021antibody}\footnote{Certain commercial equipment, instruments, software, or materials are
identified in this paper in order to specify the experimental procedure adequately. Such identification is not intended to imply recommendation or endorsement by the National Institute of Standards and Technology, nor is it intended to imply that the materials or equipment identified are necessarily the best available for the purpose.} (Section \ref{sec:1D_ex}).
We then computationally validate the convergence of  generalized prevalence estimates to the true values in mean square and illustrate a generalization to 2D data (Section \ref{sec:comp}). Finally, the discussion includes further analysis of prevalence estimation,  extensions, and limitations (Section \ref{sec:disc}).  

\section{Notation}

This work combines set and measure theory with applied diagnostics; readers will likely not be experts in both. In order for the ideas of this paper to be readily implemented by diagnostics experts and the applications understood by mathematicians, we provide baseline terminology from both fields.

\subsection{ Definitions from applied diagnostics}

\begin{itemize}

\item The na{\"i}ve class comprises individuals that have not been previously infected or vaccinated. In a binary classification, such samples are often referred to as `negative'.

\item The previously infected class comprises individuals with a prior infection but who are unvaccinated. In a binary classification, such samples are often referred to as `positive'.

\item The vaccinated class comprises individuals who have been inoculated against a disease without a prior infection.

\item  Training data correspond to samples for which the true classes are known.  Typically, such data are used to construct conditional probability models.

\item Test data correspond to samples for which the true classes are unknown or assumed to be unknown for validation purposes. Typically, a classification procedure is applied to such data.

\item  Generalized prevalence is the relative fraction of samples in a population that belong to each class. 

\end{itemize}

\subsection{Definitions from measure theory}

\begin{itemize}

\item A set is a collection of objects, e.g.\ measurement values. A domain is a set in a continuous measurement space; see Figure  \ref{fig:phase} for an example.

\item The symbol $\mathbb{R}$ denotes the set of all real numbers. The symbol $\mathbb{R}^m$  denotes the real coordinate space of dimension $m$ consisting of all real-valued vectors of length $m$.

\item  The symbol $\in$ indicates set inclusion. The expression $\bm{r} \in A$ means $\bm{r}$ is in set $A$.

\item The symbol $\subset$ denotes the subset relationship of two sets. The expression $A \subset B$ means that all elements in $A$ are contained in $B$.

\item The use of a superscript $C$ denotes the complement of a set. The set $D^C$ contains all elements in the measurement space not in $D$.

\item  The symbol $\emptyset$ denotes the empty set, which contains no elements.

\item  The operator $\cup$ denotes the union of two sets. The set $C = A \cup B$ contains all elements in either $A$ or $B$ or both.

\item The operator $\cap$ denotes the intersection of two sets. The set $C = A \cap B$ contains all elements in both $A$ and $B$.

\item  The operator $\setminus$ denotes the set difference. The set $C = A \setminus B$ contains all objects in $A$ that are not also in $B$. 
An equivalent interpretation is that $A \setminus B$ is the result of removing the common elements of $A$ and $B$ from $A$.

\item  The notation $A = \{\bm{r} : *\}$ defines the set $A$ as all $\bm{r}$ that satisfy condition $*$.

\end{itemize}

\subsection{Notation specific to this paper}

\begin{itemize}

\item The set $\Omega$ denotes the entire measurement space.

\item The label $C_j$ refers to the $j$th class. 

\item The  generalized prevalence for class $C_j$ is denoted by $q_j$.

\item The set $D_j$ denotes a domain corresponding to $C_j$.

\item The use of a superscript $\star$ denotes an optimal quantity. For example, $D_j^{\star}$ could be an optimal classification domain corresponding to class $C_j$.

\end{itemize}

\section{Generalized prevalence estimation and multiclass classification}
\label{sec:classif}

Prevalence estimation and classification rely on the same framework of antibody measurements. For each individual or sample, we represent corresponding measurements as a vector $\bm{r} = (r_{1}, \ldots r_{m} ) \in \Omega \subset \mathbb{R}^m$.  Here, $\bm{r}$ could denote $m$ antibody types targeting different parts of a virus as measured in median fluorescence intensity (MFI). Let $P_j(\bm{r})$ describe the probability that a sample from class $C_j$ yields measurement value $\bm{r}$. 
 These conditional probability density functions (PDFs) are assumed known in this section; their construction is considered for example serological data in Section \ref{sec:pdfs}.

 The generalized  prevalence $q_j$ is the relative fraction of the population corresponding to class $C_j$. In what follows we assume there are $n$ classes. The generalized prevalences must sum to 1, which implies
\begin{subequations}
\begin{eqnarray}
\sum_{j = 1}^n q_j = 1,
\label{eq:prev_cond_1} \\
q_k = 1 - \sum_{\substack{j = 1 \\ j \neq k}}^n q_j, \quad \text{ for } k \in \{1, \ldots, n\}.
\label{eq:prev_cond_2}
\end{eqnarray}
\label{eq:q_relation}
\end{subequations}
The probability density $Q(\bm{r})$ of a measurement $\bm{r}$ for a test sample is given by
\begin{equation}
Q(\bm{r}) = \sum_{j = 1}^n q_j P_j(\bm{r}).
\end{equation}
The product $q_j P_j(\bm{r})$ is the probability that a random sample both belongs to class $C_j$ and has measurement value $\bm{r}$;
thus, the expression for $Q$ is an instance of the law of total probability. 
This quantity plays an important role in prevalence estimation and classification.

\subsection{Generalized prevalence estimation}
\label{sec:pool}
\label{sec:prev}

To demonstrate the importance of prevalence in diagnostic classification, consider the United States population's SARS-CoV-2 antibody response. In early 2020, most samples should have been classified as na{\"i}ve because the disease prevalence was small. By February 2022, 
the disease prevalence was estimated at 57.7 \% \citep{clarke2022seroprevalence}; a significant fraction of samples should have been classified as previously infected. Crucially, \textit{the same measurement value may be classified differently depending on the disease prevalence}. This example shows that prevalence plays an integral role in classifying diagnostic tests and should be estimated before classification of test data.   
We address this need by designing unbiased estimators for the prevalences $\{q_j\}$.

For $n$ classes, consider a partition $\{D_j\}$
that separates the measurement space $\Omega$ into $n$ nonempty domains. It is important to note that these $D_j$ are not classification domains.
Define
\begin{equation}
Q_{j} = \int_{D_j} Q(\bm{r}) d \bm{r} = \int_{D_j}  \sum_{k = 1}^n q_{k} P_k(\bm{r}) d \bm{r} = \sum_{k = 1}^n q_{k} \int_{D_j} P_k(\bm{r}) d \bm{r} = \sum_{k = 1}^n  P_{j,k}q_{k},
\end{equation}
 where 
\begin{equation}
P_{j,k} = \int_{D_j} P_k(\bm{r}) d \bm{r}.
\end{equation}
 Writing this as a linear system yields
\begin{equation}
\left[ \begin{array}{c}
Q_1 \\ \vdots \\ Q_n
\end{array} \right] = \left[ \begin{array}{ccc}
P_{1,1} &  \ldots & P_{1,n} \\
\vdots & \ddots & \vdots \\
P_{n,1} & \ldots & P_{n,n}
\end{array} \right] \left[ \begin{array}{c}
q_1 \\ \vdots \\ q_n
\end{array} \right].
\label{eq:q_full}
\end{equation}
Taking $k = n$ in (\ref{eq:prev_cond_2}) implies
\begin{equation}
\left[ \begin{array}{c}
Q_1 \\ \vdots \\ Q_{n-1} 
\end{array} \right]  - \left[ \begin{array}{c}
P_{1,n} \\ \vdots \\  P_{n-1, n}
\end{array} \right] = \left( \left[ \begin{array}{ccc}
P_{1,1} &  \ldots & P_{1, n-1}  \\
\vdots & \ddots & \vdots \\
P_{n-1,1} & \ldots & P_{n-1, n-1} 
\end{array} \right] - \left[ \begin{array}{c}
P_{1,n}  \\ \vdots \\   P_{n-1,n} 
\end{array} \right] \underbrace{[1, \ldots, 1]}_{n-1} \right) \left[ \begin{array}{c}
q_1 \\ \vdots \\ q_{n-1}
\end{array} \right].
\end{equation}
This yields the prevalences $\bm{q}$ as the solution to the system 
\begin{subequations}
\begin{eqnarray}
\bm{q} = (\bm{P} - \bm{P_n})^{-1} \left( \bm{\overline{Q}} - \overline{\bm{P_n}} \right),
\label{eq:prev_est_m} \\
%\sum_{k = 1}^n  q_k = 1, \quad 
q_k \geq 0 \quad \text{ for } k = 1, 2, \ldots, n-1,
\label{eq:prev_est_const}
\end{eqnarray}
\label{eq:prev_est_eqn}
\end{subequations}
\noindent where $\bm{q}$ is the vector of length $n-1$ whose $j$th entry is $q_j$, $\bm{P}$ is the $(n-1) \times (n-1)$ matrix whose $(i,j)$th entry is $P_{i, j}$, $\bm{P_n}$ is the $(n-1) \times (n-1)$ matrix whose $(i,j)$th entry is $P_{i, n}$, $\bm{\overline{Q}}$ is the vector of length $n-1$ whose $j$th entry is $Q_j$, and $\overline{\bm{P_n}}$ is the vector of length $n-1$ whose $j$th entry is $P_{j, n}$. The last prevalence $q_n$ is found via  (\ref{eq:prev_cond_2}) with $k = n$.  We assume that the inverse of the matrix $\bm{P} - \bm{P_n}$ exists; Section \ref{sec:disc_lim_prev} further discusses the matrices $\bm{P}$ and $\bm{P} - \bm{P_n}$.

To estimate the generalized prevalences, estimate the $Q_j$ by $\hat{Q}_j$, where 
\begin{equation}
 Q_j \approx \hat{Q}_j = \frac{1}{S} \sum_{i = 1}^S \mathbb{I}( \bm{r}_i \in D_j).
 \label{eq:Q_approx}
 \end{equation} 
 Here, $S$ is the total number of samples and $\mathbb{I}$ denotes the indicator function. Substituting $\hat{Q}_j$ for $Q_j$ in (\ref{eq:prev_est_eqn}) yields an estimate $\hat{q}_k$ for $q_k$. When the PDFs $P_j(\bm{r})$ are known and $(\bm{P} - \bm{P_n})^{-1}$ exists, these estimates $\hat{q}_k$ are unbiased, i.e., $E[\hat{q}_k] = q_k$. This follows directly from the fact that $\hat{Q}_j$ is a Monte Carlo estimator of $Q_j$. Further, the  generalized prevalence estimates converge to the true values in mean square as the number of samples is increased \citep{caflisch1998monte}. This is illustrated in Section \ref{sec:conv_prev_est}.

We note that the  generalized prevalence estimates are not unique due to the arbitrariness of the $\{D_j\}$. 
However, the non-uniqueness allows us to select any reasonable partition over which to find the estimators $\{ \hat{Q}_j\}$. This is discussed further in Section \ref{sec:disc_lim_prev}.

\subsection{Optimal classification}
\label{sec:opt_class}

Our task is to define a partition $\{D_j\}$ (not necessarily the same as for prevalence estimation) of the measurement space $\Omega$ such that each domain corresponds to one and only one class $C_j$. A measurement $\bm{r}$ is assigned to class $j$ if $\bm{r} \in D_j$. We require
\begin{subequations}
\begin{eqnarray}
 \mu_{j}\left(\bigcup_{k = 1}^n D_k \right) = 1 \quad \forall j \in \{1, \ldots, n\},
 \label{eq:req2} \\
 \mu_{\ell}(D_j \cap D_k) = 0 \text{ for } j \neq k, \quad \forall \ell = 1, 2, \ldots, n
 \label{eq:req1}
\end{eqnarray}
\label{eq:req}
\end{subequations}
where $ \mu_{\ell}(X) = \int_X P_{\ell}(\bm{r}) d\bm{r}$. Here, (\ref{eq:req2}) ensures that any sample can be classified and (\ref{eq:req1}) enforces single-label classification (up to sets of measure zero). To identify an optimal partition $\{D_j^{\star}\}$, we construct the loss function
\begin{equation}
\mathscr{L}(D_1, \ldots, D_n) = \sum_{j = 1}^n q_j \int_{\Omega \setminus D_j} P_j(\bm{r}) d\bm{r}.
\label{eq:loss}
\end{equation}
Here, (\ref{eq:loss}) is the  generalized prevalence-weighted convex combination of false classification rates as a function of the domains $D_j$. Intuitively, we expect that a sample with measurement $\bm{r}$ should be assigned to the class domain $D_j$ to which it has the highest probability of belonging; that is, the highest value of $q_j P_j(\bm{r})$ for all $j$. Accordingly, the loss function (\ref{eq:loss}) penalizes misclassified measurements $\bm{r}$ with high probability values.

To address situations in which a measurement has an equal highest probability of belonging to two or more classes, we introduce the following set for each class $C_j$: 
\begin{equation}
\mathscr{E}_j =  \bigcup_{\substack{k = 1 \\ k \neq j}}^n \{ \bm{r}:  q_k P_k(\bm{r}) =  q_j P_j(\bm{r}) = \max_i q_i P_i(\bm{r})  \} .
\end{equation}
In most practical implementations all $ \mathscr{E}_j$ have measure zero, and 
the domains
\begin{equation}
D_j^{\star} = \{\bm{r} : q_j P_j(\bm{r}) > q_k P_k (\bm{r}) \text{ for } k \neq j\}
\label{eq:opt_d}
\end{equation} 
minimize the loss function $\mathscr{L}$ up to sets of measure zero.  The proof is shown in \ref{sec:app_a} and involves a straightforward application of set theory; see also \cite{williams2006gaussian} for similar ideas.

If $\mathscr{E}_j$ has nonzero measure, randomly assigning a measurement in $\mathscr{E}_j$ to one of the classes to which it has equal maximal probability of belonging does not effect the loss $\mathscr{L}$. 
In this case, the optimal domains are generalized to
\begin{equation}
D_j^{\star} = \{\bm{r} : q_j P_j(\bm{r}) > q_k P_k (\bm{r}) \text{ for } k \neq j\} \cup Z_{\mathscr{E}_j},
\end{equation}
where $Z_{\mathscr{E}_j}$ is an element of a partition of $\mathscr{E}_j$ that we define iteratively as follows.
\begin{subequations}
\begin{eqnarray}
 Z_{\mathscr{E}_1} = \mathscr{E}_1, \\
 Z_{\mathscr{E}_k} = \mathscr{E}_k \mathbin{\big \backslash} \bigcup_{j = 1}^{k-1} \mathscr{E}_j, \quad k \in \{ 2, \ldots, n\} .
 \end{eqnarray}
 \end{subequations}
 This ensures that no measurement in a set $\mathscr{E}_j$ is assigned to more than one optimal domain. Note that by construction, $Z_{\mathscr{E}_n}$ is empty.
 
\begin{figure}[t]
\centering
\includegraphics[scale=.5]{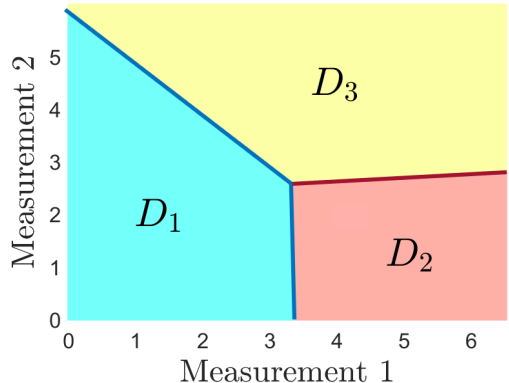}
\caption{Illustration of classification domains $D_1, D_2$, and $D_3$ in which the sets of equal probability of a measurement belonging to two or more classes are shown as lines separating the optimal regions.}
\label{fig:phase}
\end{figure}

 Figure \ref{fig:phase} shows a 2D conceptual illustration of $\{ \mathscr{E}_j\}$, which are the lines delineating the optimal regions. In 2D, line segments have Lebesgue measure zero. Thus, classification follows  (\ref{eq:opt_d}).  Note that the ``multipoint" at which the lines meet has equal probability of belonging to all three classes. We discuss this further in Section \ref{sec:loc_acc}. 

\section{Example applied to SARS-CoV-2 antibody data with three classes}
\label{sec:1D_ex}

To demonstrate the concepts developed in Section \ref{sec:classif}, we apply our methods to serological data with three classes. Publicly available data sets associated with \cite{ainsworth2020performance} and \cite{wei2021antibody} provide previously infected, na{\"i}ve, and vaccinated antibody measurements. The vaccine data \citep{wei2021antibody} are recorded for individuals that were innoculated with one of two vaccines. We refer to these as Vaccine A and Vaccine B and analyze the populations separately and together.  The studies provide SARS-CoV-2 anti-spike immunoglobulin (IgG) antibody measurements; see \ref{sec:app_MFI}  for measurement details. We use one-dimensional (1D) data to illustrate a straightforward multiclass example; Section \ref{sec:2D} demonstrates that our analysis holds for higher measurement dimensions. 

All data are transformed to a logarithmic scale as follows:
\begin{equation}
r = \log_2(\tilde{r} + 2) - 1.
\label{eq:log_transform}
\end{equation}
Here, $\tilde{r}$ and $r$ represent the original and log-transformed values; $\tilde{r}$ has units of ng/mL and $r$ is nondimensional.
This transformation puts the data on the scale of bits and allows for better viewing of measurements that range over several decades of MFI values in the original units. \cite{wei2021antibody} truncated vaccinated samples, with lower and upper transformed limits of 1 and roughly 8.

\begin{figure}[t]
\centering
\subfloat[][Vaccine A and Vaccine B split]{\includegraphics[scale=.5]{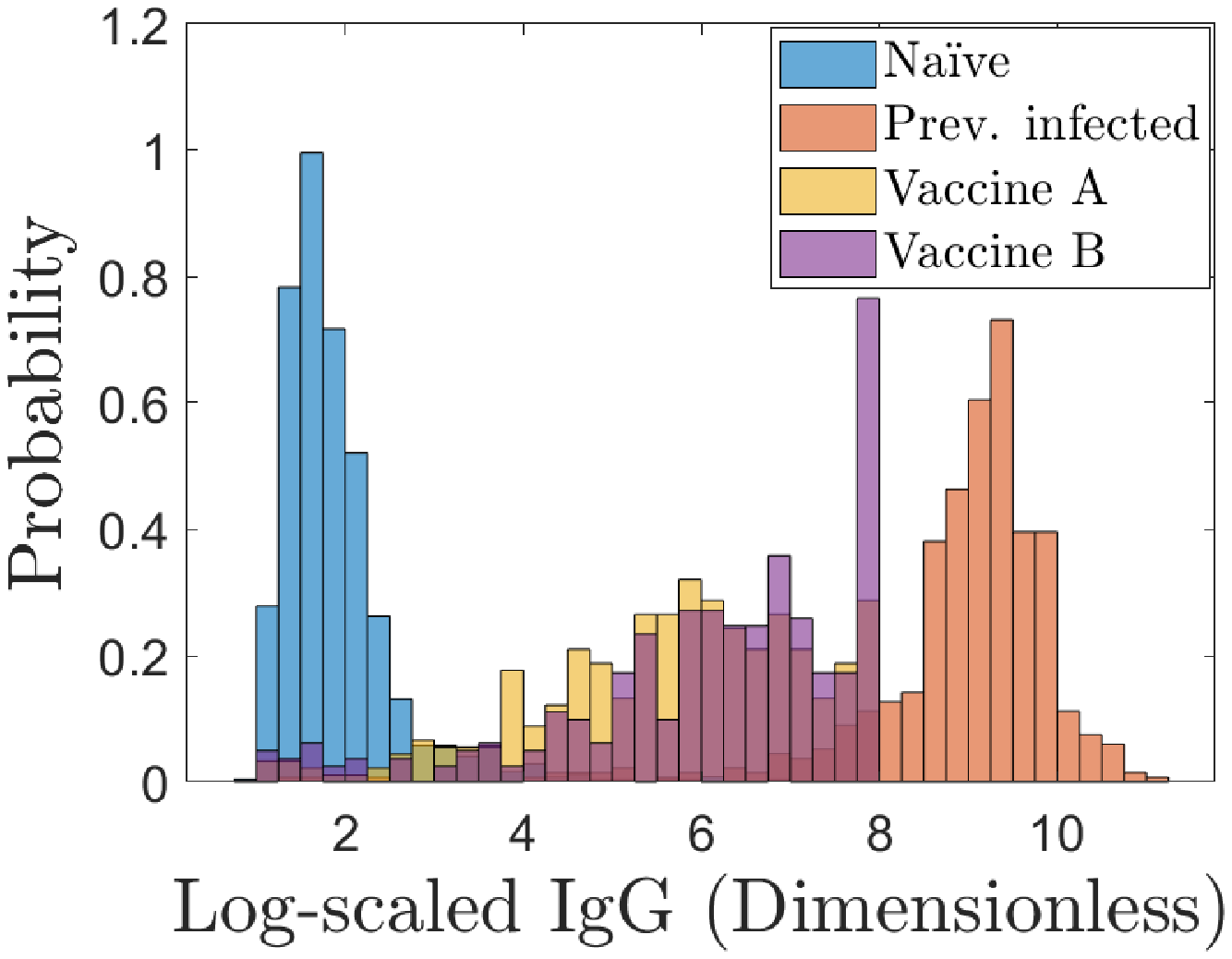}
\label{fig:1D_hist_split}}
\subfloat[][Vaccine A and B combined]{\includegraphics[scale=.5]{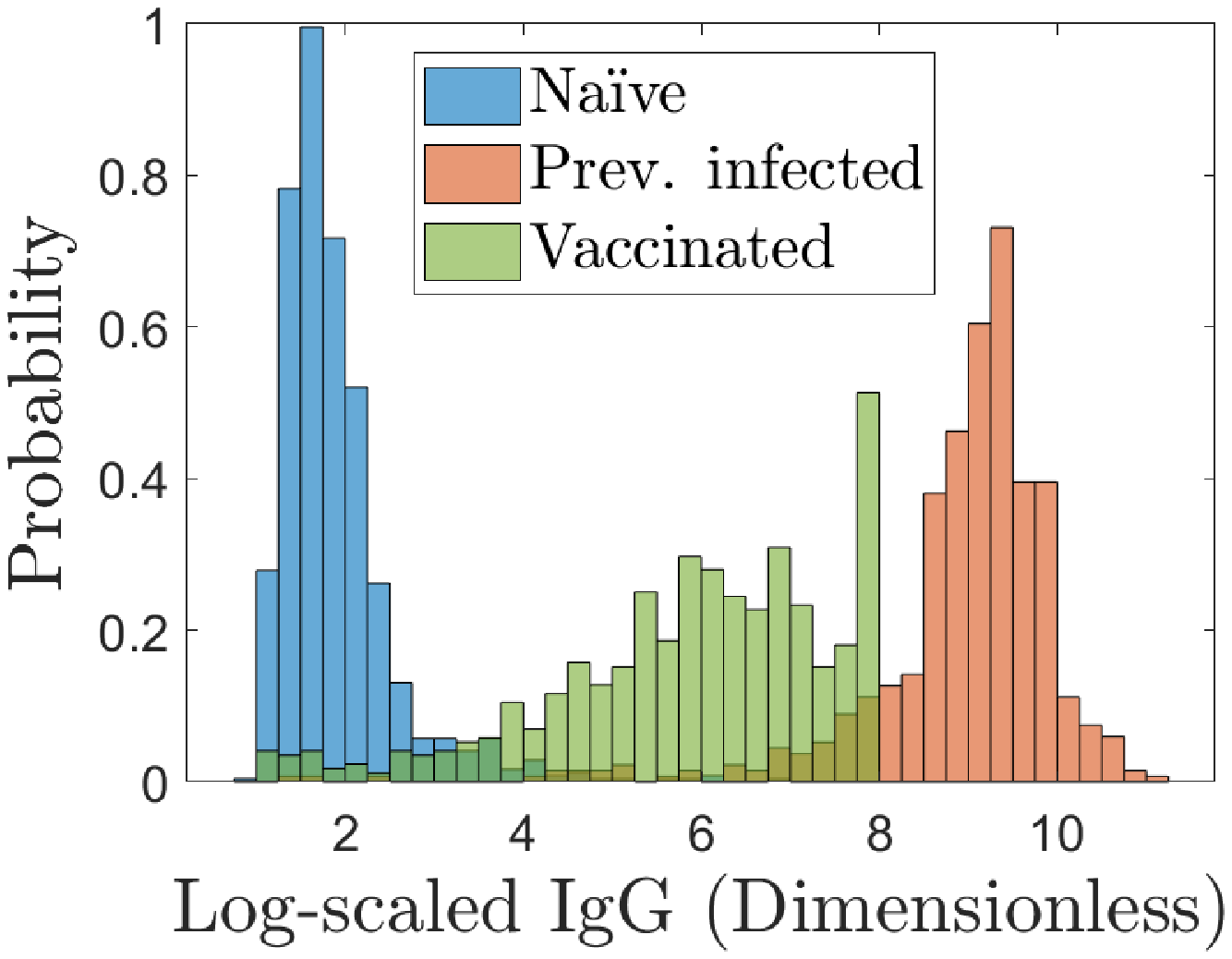}
\label{fig:1D_hist_all}}
\caption{Histograms of previously infected, na{\"i}ve, and vaccinated data from Ainsworth et. al (2020) and Wei et al. (2021). 
}
\label{fig:1D_hist_data}
\end{figure}

Figure \ref{fig:1D_hist_data} shows a histogram of the data with the vaccinated category split by vaccine manufacturer (Figure \ref{fig:1D_hist_split}) and combined (Figure \ref{fig:1D_hist_all}). 
Previously infected samples have the largest spike IgG antibody levels and na{\"i}ve samples the smallest; the vaccinated class falls in the middle. The vaccinated class overlaps with some na{\"i}ve and previously infected samples. Due to the truncation of vaccinated measurements, the corresponding right-most histogram bin contains many samples.
We separate the data into randomly generated training (80 \% of samples) and test (20 \%) populations. 

\subsection{Conditional probability distributions}
\label{sec:pdfs}

We fit probability distributions to the training data to model the na{\"i}ve, previously infected, and vaccinated antibody responses. For our purposes, we assume these are distinct classes; a sample belongs to one and only one of the three categories.
To construct the conditional PDF for each population, we select a parameterized model that empirically characterizes the shape and spread of the samples.
We determine parameters separately for the three training populations by maximum likelihood estimation (MLE). 
The na{\"i}ve training population is fit to a Burr distribution
\begin{equation}
N(r) = \frac{ck}{\lambda} \left( \frac{r}{\lambda} \right)^{c-1} \left[ 1 +\left( \frac{r}{\lambda}\right)^c \right]^{-k-1},
\label{eq:N_eqn}
\end{equation}
which describes a right-skewed sample population.

The previously infected  training population is fit to a stable distribution described by characteristic function
\begin{equation}
\phi(r) = \exp\left\{i r \delta - |\gamma r|^{\alpha} \left[1 + \frac{2i}{\pi} \beta \text{sgn}(r) \log(\gamma r)\right] \right\}
\label{eq:P_eqn}
\end{equation}
for $\gamma \neq 1$. 
Here, $i$ is the imaginary unit and sgn is the sign function, which returns +1, -1, or 0. This distribution describes a left-skewed sample population. 

We fit the vaccinated training populations to an extreme-value distribution after observing the mostly symmetric shape of the data with a spike at the right truncation limit:
\begin{equation}
V(r) = \frac{1}{\sigma} \exp \l \frac{r - \mu}{\sigma} \r \exp \left[ - \exp \l \frac{r - \mu}{\sigma} \r \right].
\label{eq:V_eqn}
\end{equation}
We apply data censoring to better fit the truncated data; this is described in \ref{sec:app_trunc}.

\begin{figure}[h]
\centering
\includegraphics[scale=.5]{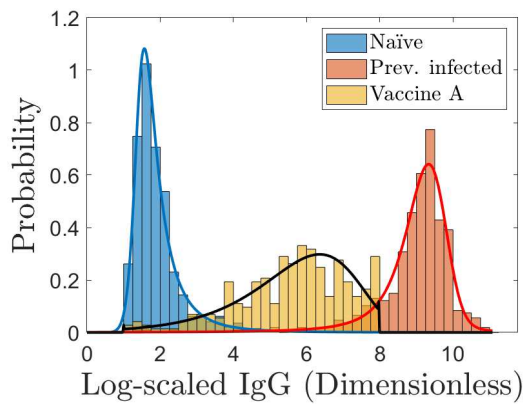}
\caption{Conditional PDFs for the three na{\"i}ve, previously infected, and vaccinated classes trained on the training data for a Vaccine A visualization of the vaccinated class. See Supplemental Figure S1 for the PDFs of the Vaccine B and combined visualizations. }
\label{fig:pdf}
\end{figure}

The analysis in this section is identical for all three visualizations of the vaccinated class. In what follows, we report all results but show only the Vaccine A figures as examples. Corresponding figures for the Vaccine B and combined visualizations of the vaccine class are left to the Supplemental Data. 

Figure \ref{fig:pdf} shows the conditional PDFs, represented as continuous curves, trained on the three-class training data with a Vaccine A vaccinated class. The blue, red, and black curves correspond to the na{\"i}ve, previously infected, and vaccinated models.
The effect of truncating the data at the upper limit is visible in the right-most bin of the vaccinated class histogram; this is accounted for by the data-censoring. As a result, the vaccinated class PDF exhibits spikes at the upper and lower truncation values. This spike is an artifact of the original data collection process and not a typical problem.

\subsection{Generalized prevalence estimation}

\begin{figure}[h]
\centering
\includegraphics[scale=.5]{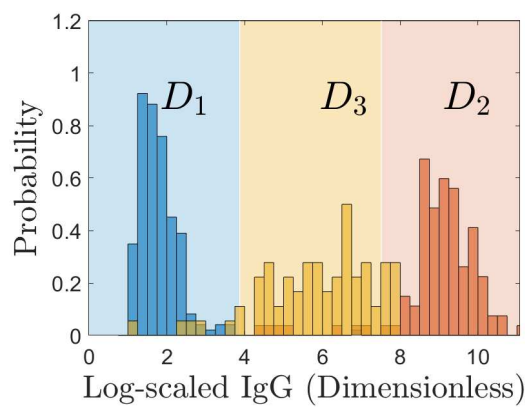}
\caption{Test data $k$-means partitioning with a Vaccine A vaccinated class for  generalized prevalence estimates. We use $k = 3$ classes; the clustered domains are labeled as $D_1, D_2$, and $D_3$. See Supplemental Figure S2 for the partitions of the Vaccine B and combined visualizations of the vaccinated class.}
\label{fig:part}
\end{figure}

Recall that prevalence estimation of test data requires a partition that separates the measurement space $\Omega$ into  $n$ nonempty domains. Here, the number of classes is $n = 3$. We create a partition using $k$-means clustering with $k = 3$, which assigns each measurement to the cluster with the closest mean. Figure \ref{fig:part} shows the partition for our test data set with a Vaccine A vaccinated class. The clustering separates the three populations reasonably well; see Section \ref{sec:disc_lim_prev} for the importance of this statement. The partition need not perfectly separate the data by class to estimate prevalences with high accuracy.
We estimate  generalized prevalences for the test data via (\ref{eq:prev_est_eqn}) and record true and estimated values in Table \ref{table:prev_est}. 

\begin{table}[h]
\centering
\begin{tabular}{|l|r|rrr|rrr|r|}
\hline
 & \multicolumn{1}{l|}{} & \multicolumn{3}{c|}{\textbf{\begin{tabular}[c]{@{}c@{}}Vaccine data set\\ Estimated (true)  generalized prevalence\end{tabular}}} & \multicolumn{3}{c|}{\textbf{\begin{tabular}[c]{@{}c@{}}Errors\\ (\%)\end{tabular}}} & \multicolumn{1}{c|}{\textbf{\begin{tabular}[c]{@{}c@{}}Avg.\\ errors\\ (\%)\end{tabular}}} \\ \hline
 & \multicolumn{1}{l|}{\textbf{}} & \multicolumn{1}{c|}{\textit{A}} & \multicolumn{1}{c|}{\textit{B}} & \multicolumn{1}{c|}{\textit{All}} & \multicolumn{1}{c|}{\textit{A}} & \multicolumn{1}{c|}{\textit{B}} & \multicolumn{1}{c|}{\textit{All}} &  \\ \hline
\multirow{3}{*}{\textbf{Class}} & $N$ & \multicolumn{1}{r|}{0.523 (0.521)} & \multicolumn{1}{r|}{0.538 (0.531)} & \multicolumn{1}{r|}{0.445 (0.444)}  & \multicolumn{1}{r|}{0.377}& \multicolumn{1}{r|}{1.02} &  {0.223}  & 0.540 \\ \cline{2-9} 
 & $P$ & \multicolumn{1}{r|}{0.300 (0.286)}  & \multicolumn{1}{r|}{0.313 (0.292)} &  {0.285 (0.244)} & \multicolumn{1}{r|}{4.69} & \multicolumn{1}{r|}{7.10} & {17.0} & 9.93 \\ \cline{2-9} 
 & $V$ & \multicolumn{1}{r|}{0.177 (0.193)} & \multicolumn{1}{r|}{0.149 (0.177)} & {0.270 (0.312)} & \multicolumn{1}{r|}{7.99} & \multicolumn{1}{r|}{15.0} & {13.6} & 12.2 \\ \hline
\textbf{Avg.} &  & \multicolumn{1}{r|}{} & \multicolumn{1}{r|}{} &  & \multicolumn{1}{r|}{4.74}  & \multicolumn{1}{r|}{7.67} & {10.3} & 7.55 \\ \hline
\end{tabular}
\caption{ Estimated and true  generalized prevalences for the test data na{\"i}ve (N), previously infected (P), and vaccinated (V) classes.  Vaccine A (A) and Vaccine B (B) are considered separately and together (All).}
\label{table:prev_est}
\end{table}

\subsection{Optimal classification}

\begin{figure}[h]
\centering
\subfloat[][Training data]{
\includegraphics[scale=.5]{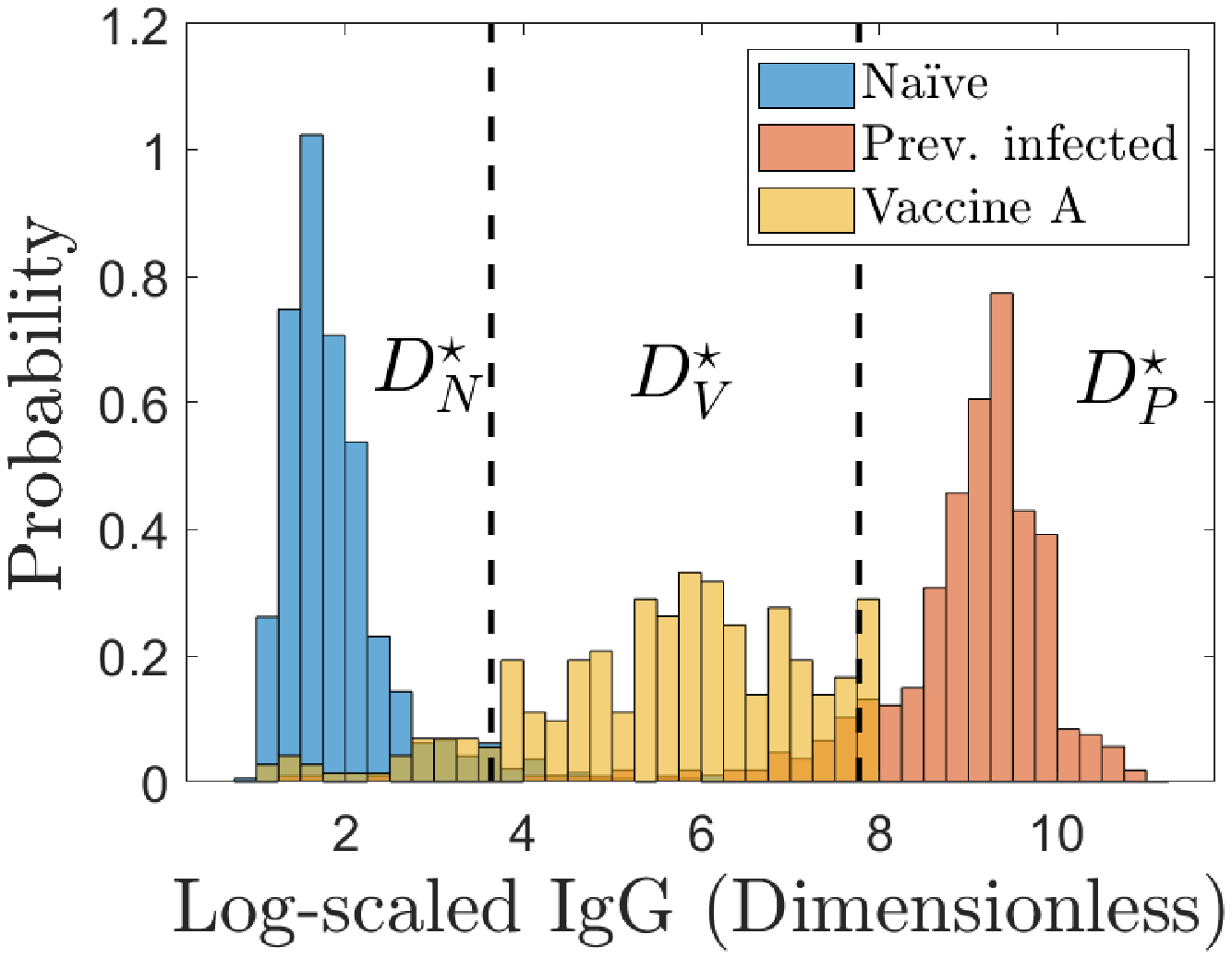}\label{fig:train}}
\subfloat[][Test data]{\includegraphics[scale=.5]{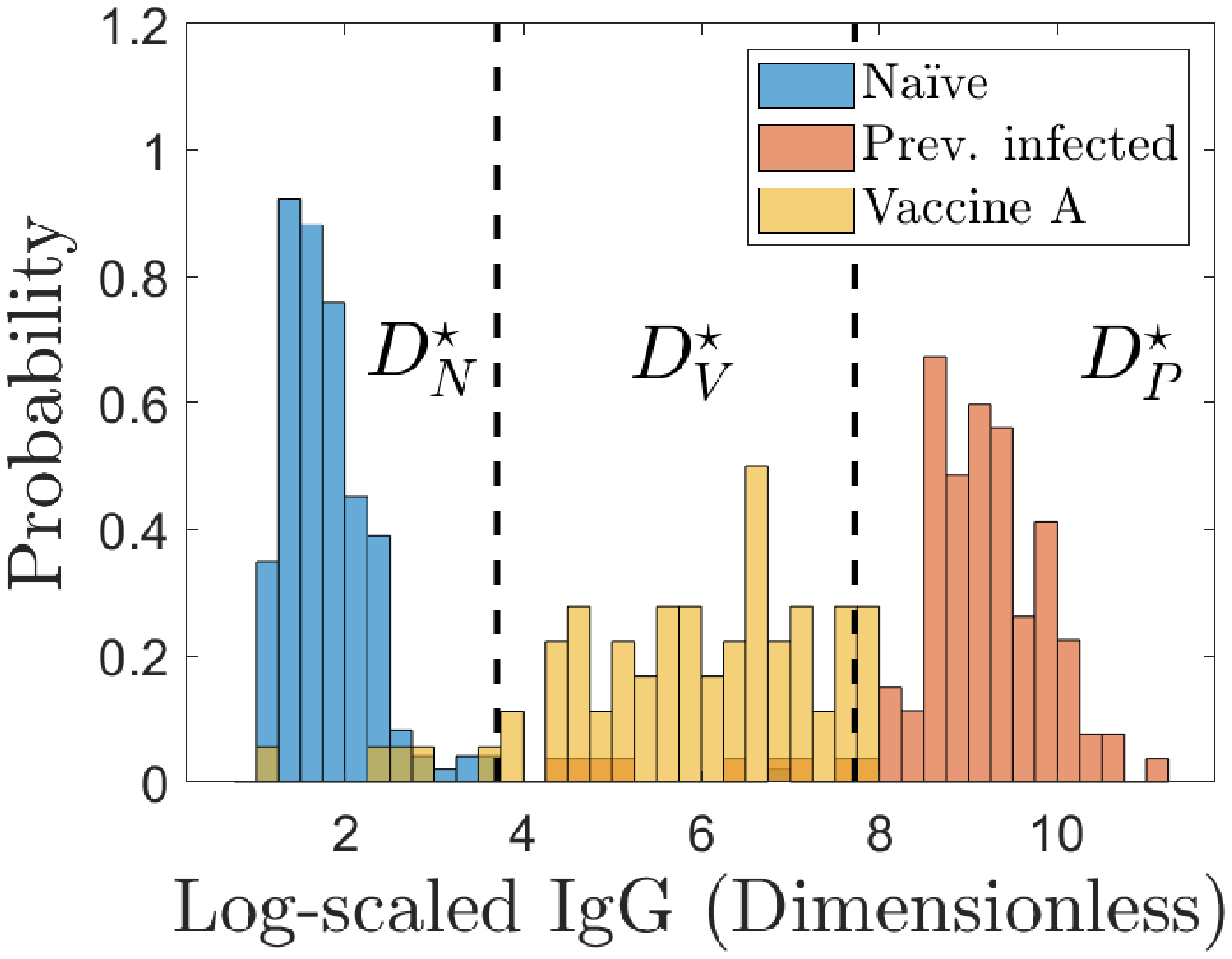}\label{fig:test_opt}}
\caption{Training (a) and test (b) data with a Vaccine A vaccinated class with optimal decision thresholds using a known prevalence.  
Vertical dashed lines indicate the optimal decision boundaries. The optimal na{\"i}ve, vaccinated, and previously infected domains are labeled $D_N^{\star}$, $D_V^{\star}$, and $D_P^{\star}$. See Supplemental Figures S3 and S4 for the optimal domains for Vaccine B and combined visualizations of the vaccinated class.
}
\label{fig:train_test}
\end{figure}

We classify the training data using known generalized prevalences via  (\ref{eq:opt_d}).
Figure \ref{fig:train} shows the optimal domains, labeled $D_N^{\star}$, $D_V^{\star}$, and $D_P^{\star}$, for a Vaccine A vaccinated class. 
 For this 1D example with three classes, the optimal classification domain boundaries can be represented by upper and lower threshold levels. Samples with measurements below the smaller level are classified as na{\"i}ve, samples with measurements between the thresholds as vaccinated, and samples with measurements above the larger level as previously infected. All three populations have overlapping PDFs, which reduces classification accuracy.
 
 Accurate classification of test data is possible with reasonably close prevalence estimates. We classify the test data using estimated generalized prevalences and display the optimal classification domains for a Vaccine A vaccinated class in Figure \ref{fig:test_opt}.

\begin{table}[h]
\centering
\begin{tabular}{|c|r|r|}
\hline
\multicolumn{1}{|l|}{} & \multicolumn{1}{c|}{\textbf{\begin{tabular}[c]{@{}c@{}}Training\\ classification \\ error (\%)\end{tabular}}} & \multicolumn{1}{c|}{\textbf{\begin{tabular}[c]{@{}c@{}}Test\\ classification\\ error (\%)\end{tabular}}} \\ \hline
\textbf{Vaccine A} & 7.87  & 5.08 \\ \hline
\textbf{Vaccine B} & 7.07 &  5.46 \\ \hline
\textbf{Combined} & 7.90 & 4.78 \\ \hline
\end{tabular}
\caption{Classification errors for training data using a known prevalence and test data with an estimated prevalence. Vaccine A and Vaccine B are considered separately and together (Combined).}
\label{table:train_test_classif}
\end{table}

Training and test data classification errors are recorded in Table \ref{table:train_test_classif}. Taken over the three considerations of the vaccinated class, the average error for the training data is 7.61 \% and the same for the test data is 5.11 \%.

\section{Computational validation}
\label{sec:comp}

We numerically demonstrate two important features of our generalized prevalence estimation and multiclass optimal classification procedures. First, we show the convergence of our prevalence estimates to the true values as the number of samples is increased. Second, we present a 2D tri-class problem to show how the method generalizes to higher dimensional measurement spaces.

\subsection{Convergence of prevalence estimates}
\label{sec:conv_prev_est}

We use our probability models (\ref{eq:N_eqn})-(\ref{eq:V_eqn}) to generate synthetic data sets whose relative frequencies match the  generalized prevalences of the \cite{ainsworth2020performance} and \cite{wei2021antibody} data. 
We systematically increase the number of synthetic data points used while holding generalized prevalences fixed to study the effect of sample size on prevalence convergence. For each number of points used, we generate 1000 synthetic data sets and compute statistics on our results. 

\begin{figure}[t]
\centering
\subfloat[][1000 synthetic samples]{\includegraphics[scale=.5]{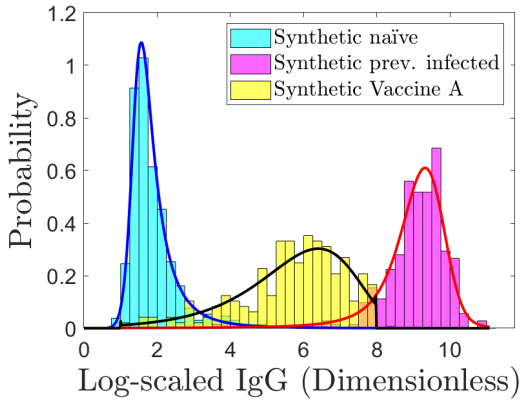}
\label{fig:syn_AZ_data}}
\subfloat[][ Boxchart of estimate statistics]{\includegraphics[scale=0.5]{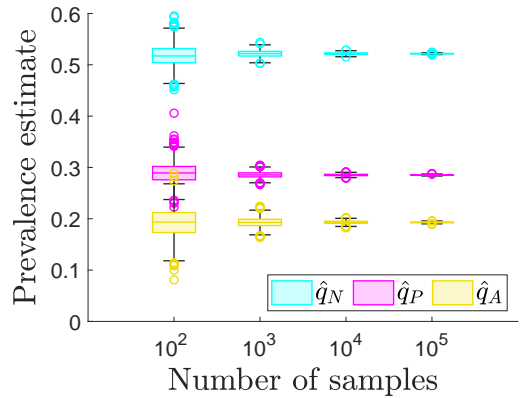} 
\label{fig:syn_AZ_boxchart}} \\
\subfloat[][ Convergence of the error in mean square]{\includegraphics[scale=0.5]{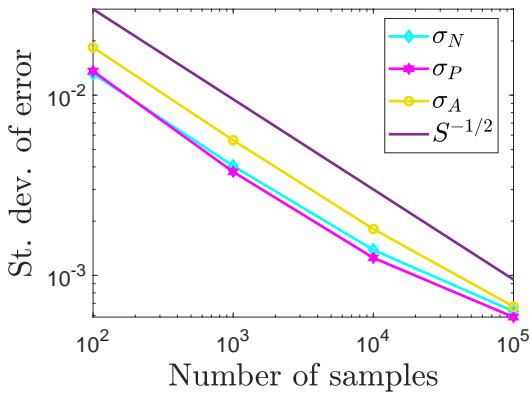} 
\label{fig:syn_AZ_conv}}
\begin{minipage}[b]{18em}
\caption{Prevalence estimation convergence for synthetic data using a Vaccine A vaccinated class (1000 simulations). 
 In (b), the boxes display the median and upper and lower quartiles as the line inside the box and its top and bottom edges. The whiskers show non-outlier maximum and minimum values; outliers vary from the median by more that 1.5 times the difference between the upper and lower quartiles, and are shown as circles. In (b) and (c), the subscripts $N$, $P$, and $A$ denote na{\"i}ve, previously infected, and Vaccine A vaccinated. The number of samples is $S$.}
\label{fig:syn_AZ}
\end{minipage}
\end{figure}

\begin{table}[t]
\centering
\begin{tabular}{|l|r|r|r|r|}
\hline
 & \multicolumn{1}{l|}{} & \multicolumn{1}{c|}{\textbf{Na{\"i}ve}} & \multicolumn{1}{c|}{\textbf{Previously infected}} & \multicolumn{1}{c|}{\textbf{Vaccine A}} \\ \hline
\textbf{\begin{tabular}[c]{@{}l@{}}True generalized\\ prevalences\end{tabular}} & \multicolumn{1}{l|}{} & 0.521 & 0.286 & 0.193 \\ \hline
\multirow{4}{*}{\textbf{Number of samples}} & $10^2$ & 0.518 $\pm$ 0.0212 & 0.289 $\pm$ 0.0211 & 0.193 $\pm$ 0.0299 \\ \cline{2-5} 
 & $10^3$ & 0.522 $\pm$ 0.0067 & 0.286 $\pm$ 0.0061 & 0.193 $\pm$ 0.0093 \\ \cline{2-5} 
 & $10^4$ & 0.522 $\pm$ 0.0021 & 0.285 $\pm$ 0.0020 & 0.193 $\pm$ 0.0030 \\ \cline{2-5} 
 & $10^5$ & 0.522 $\pm$ 0.0007 & 0.285 $\pm$ 0.0007 & 0.193 $\pm$ 0.0011 \\ \hline
\end{tabular}
\caption{ Generalized prevalence estimate means and standard deviations taken over 1000 simulations of synthetic data generated from the probability models for increasing number of samples for a Vaccine A vaccinated class.}
\label{table:prev_conv}
\end{table}

Figure \ref{fig:syn_AZ} shows our analysis for the Vaccine A vaccinated class. 
Figure \ref{fig:syn_AZ_boxchart} shows a boxchart of the statistics for using $10^2, 10^3, 10^4$, and $10^5$ samples. The estimates have more outliers and variation when few samples are used, which decreases as the number of points is increased. Even with few samples, the median  generalized prevalence estimates are close to the true  generalized prevalences. Table \ref{table:prev_conv} records the mean and standard deviations of our results. Even for only 1000 samples, our estimates agree with the true  generalized prevalences with roughly 2 \% relative error.

Figure \ref{fig:syn_AZ_conv} plots the standard deviation of the  prevalence estimate error on a log-log scale against the number of samples. 
The standard deviation should decrease with the inverse square root of the number of samples \citep{caflisch1998monte}, which is plotted for comparison. Our empirical convergence rates all agree with the theory through 10,000 samples; the rate is maintained for the Vaccine A vaccinated class.

\subsection{Generalization to higher dimensions}
\label{sec:2D}

\begin{figure}[t]
\centering
\subfloat[][Probability model contours and synthetic data]{\includegraphics[scale=.5]{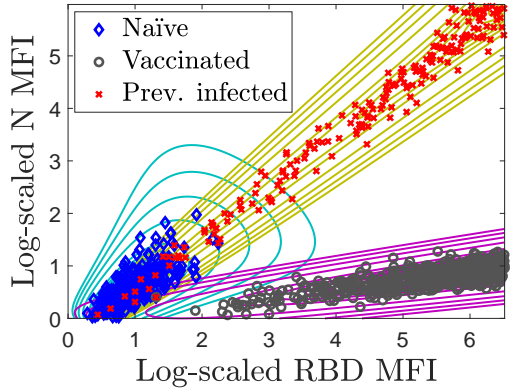}}
\subfloat[][$k$-means clustering]{\includegraphics[scale=.5]{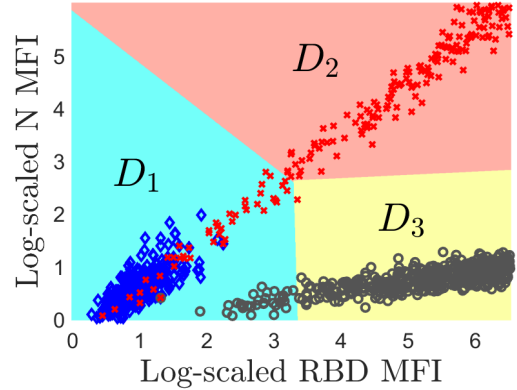}} \\
\subfloat[][Optimal classification domains]{
\includegraphics[scale=.5]{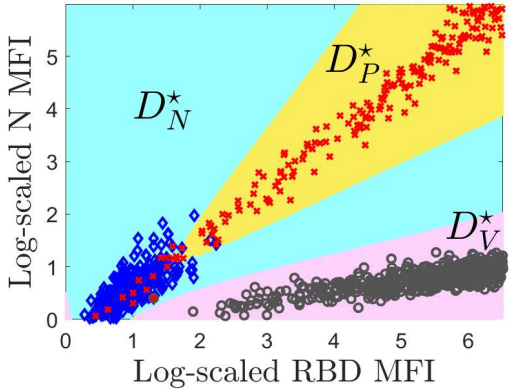}} 
\begin{minipage}[b]{18em}
\caption{(a) Level sets of conditional PDFs with example synthetic data, (b) $k$-means clustering, (c) optimal classification domains with estimated generalized prevalences. In (c), the subscripts $N$, $P$, and $V$ denote na{\"i}ve, previously infected, and vaccinated.
}
\label{fig:3_contours}
\end{minipage}
\end{figure}

We now explore a 2D synthetic numerical validation of  generalized prevalence estimation and multiclass optimal classification. See \cite{luke2022improving} for a discussion of the implications of higher-dimensional modeling on diagnostic testing accuracy. The synthetic values we use are modeled off the receptor-binding domain (RBD) and nucleocapsid (N) SARS-CoV-2 antibody targets; together these form a measurement double $\bm{r}$. Details about the models and information about the data are given in \ref{sec:app_b}. Figure \ref{fig:3_contours}a shows an example of 2D synthetic antibody measurements with na{\"i}ve, previously infected, and vaccinated classes with true  prevalences of 0.3, 0.2, and 0.5. The conditional PDFs are shown as contour lines of constant probability. We use 1000 total synthetic samples. 

\begin{figure}[t]
\centering
\subfloat[][Na{\"i}ve]{\includegraphics[scale=.5]{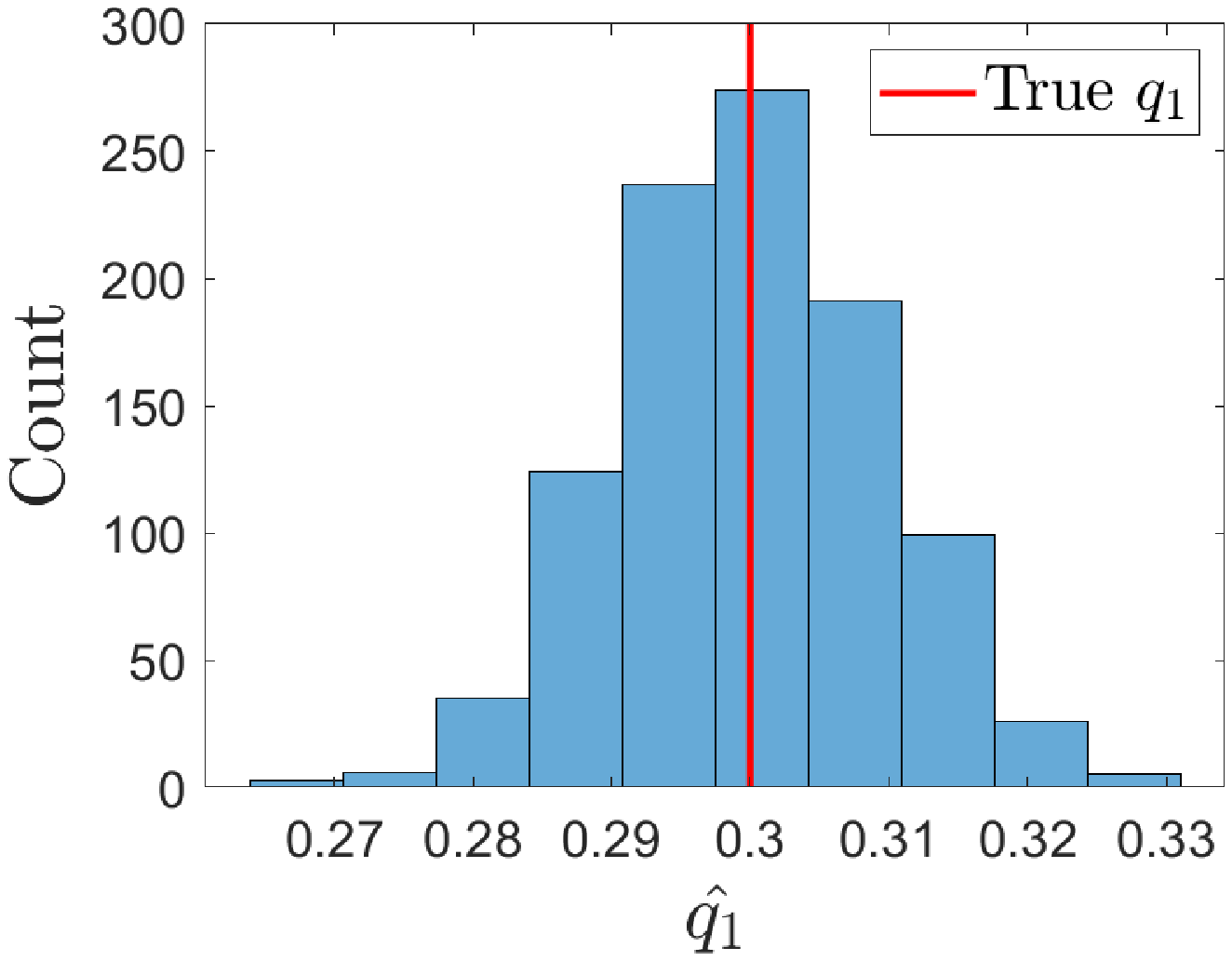}}
\subfloat[][Previously infected]{\includegraphics[scale=.5]{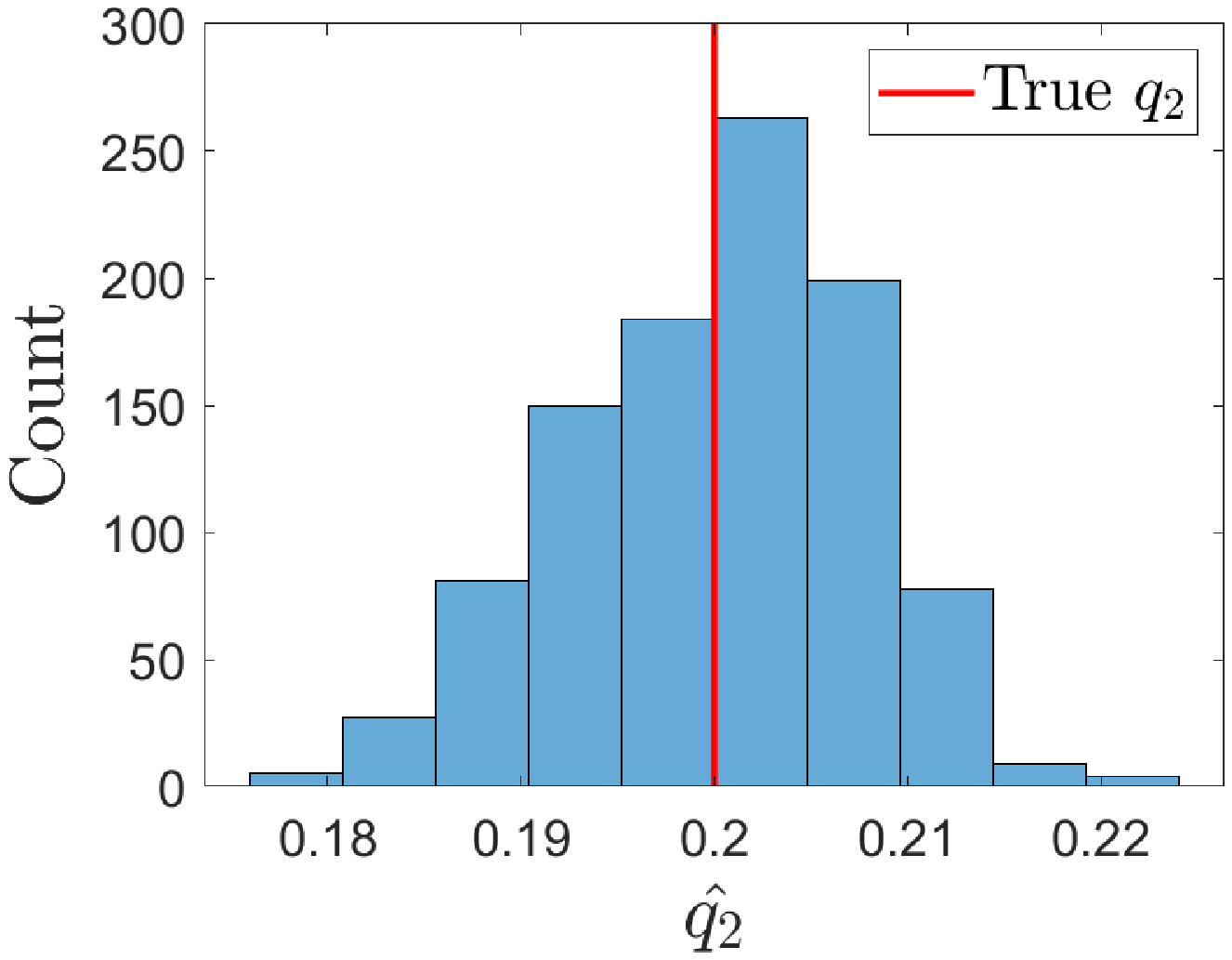}} \\
\subfloat[][Vaccinated]{\includegraphics[scale=.5]{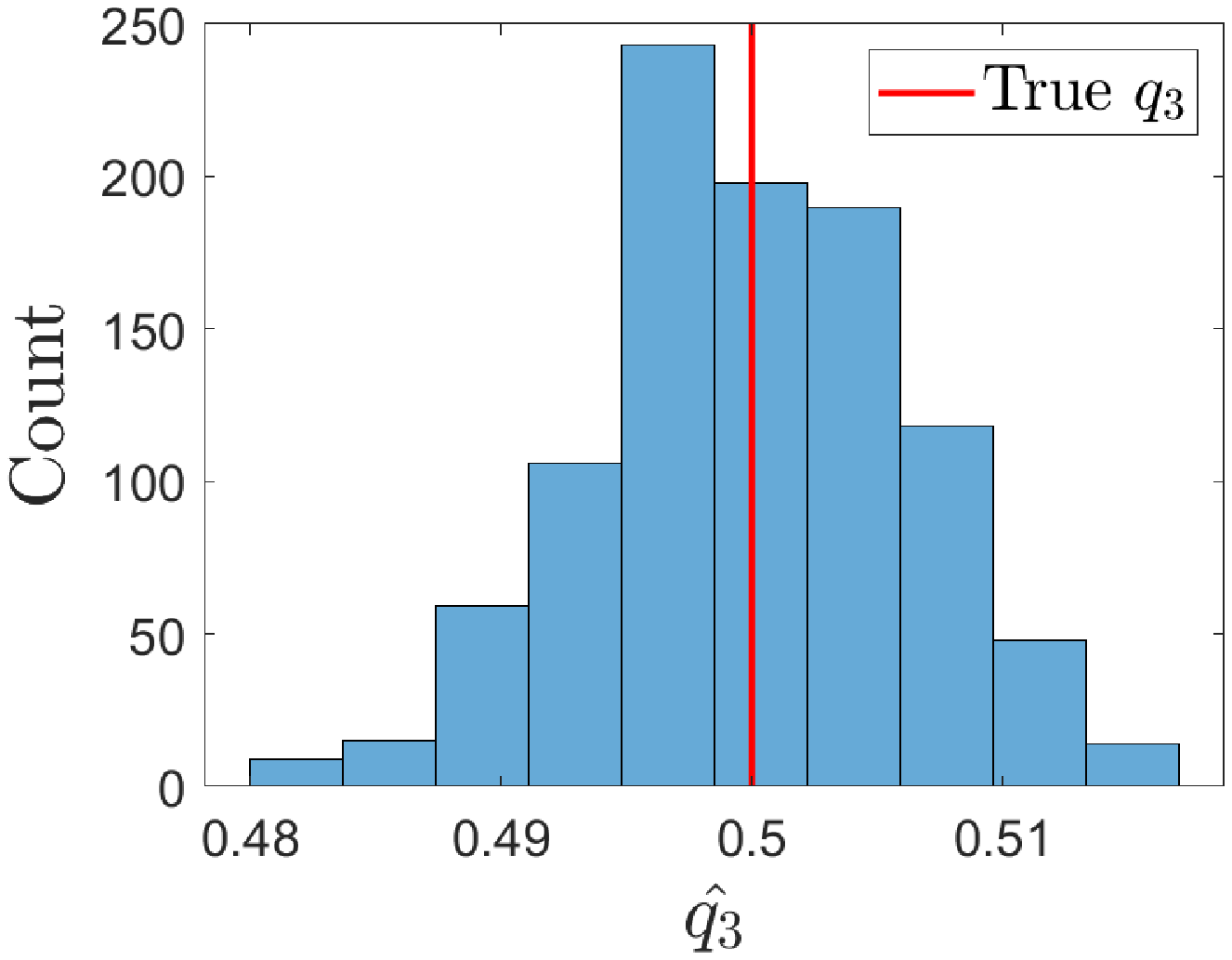}}
\begin{minipage}[b]{18em}
\caption{(a-c): Histograms of  generalized prevalence estimates from 1000 synthetic data sets. Ten bins are used for each histogram and the true prevalence is shown as a vertical red line.}
\label{fig:q_hist}
\end{minipage}
\end{figure}

To quantify uncertainty in the prevalence estimates, we randomly generate 1000 synthetic sets of samples using fixed prevalences. We then partition the measurement space via $k$-means clustering using one synthetic sample set (see Figure \ref{fig:3_contours}b), fix the partition, and use (\ref{eq:prev_est_eqn}) to generate prevalence estimates for all sets. 
The results are shown in Table \ref{table:q_stats}.  Figure \ref{fig:q_hist} shows histograms of the  generalized prevalence estimates and true values, which fall within the middle of each distribution.  We classify using these estimated prevalences via (\ref{eq:opt_d}) and find an average error of 1.58 \%. Figure \ref{fig:3_contours}c shows example optimal classification domains. The gold region is the previously infected domain, the purple is the vaccinated, and the remainder of the measurement space, colored in light blue, defines the na{\"i}ve domain. For this example,
the false classification rate is  1.8 \%. 

\begin{table}[h]
\centering
\begin{tabular}{|c|r|r|r|r|}
\hline
\multicolumn{1}{|l|}{} & \multicolumn{1}{|l|}{\textbf{True val}} & \multicolumn{1}{c|}{$\bm{\mu}$} & \multicolumn{1}{c|}{$\bm{\sigma}$} & \multicolumn{1}{c|}{\textbf{CV} $\bm{\left( \frac{\sigma}{\mu}\right)}$} \\ \hline
$\bm{q_1}$     & 0.3        & 0.300                          & $9.6 \times 10^{-3}$   & 0.0319                                                    \\ \hline
$\bm{q_2}$     & 0.2        & 0.200                          & $7.4 \times 10^{-3}$                                          & 0.0371                                                     \\ \hline
$\bm{q_3}$    & 0.5         & 0.500                           & $6.3 \times 10^{-3}$                               & 0.0126                                                     \\ \hline
\end{tabular}
\caption{ Statistics for the data shown in Figure \ref{fig:q_hist}. The true values of the prevalences are given along with the mean $\mu$, standard deviation $\sigma$, and coefficient of variation (CV) of the estimates. }
\label{table:q_stats}
\end{table}

\section{Discussion}
\label{sec:disc}

\subsection{Limiting cases of prevalence estimation and implications for assay design}
\label{sec:disc_lim_prev}

An interesting observation of our prevalence estimation scheme is that the structure of the matrix underpinning the linear system encodes information about overlap between populations. As such, the matrix potentially informs best practices for prevalence estimation. 
Further, our method extends the binary procedure of \cite{patrone2021classification} and may provide insight into the simpler setting. Here we examine limiting cases of prevalence estimation and connect characteristics of the matrix $\bm{P}$ to assay accuracy.

 We explore interpretations of equivalent definitions of singularity of the matrix $\bm{P}$. Recall that the quantity $P_{j,k}$ gives the probability density of class $k$ falling in domain $D_j$. If all elements of a row (column) of the matrix $\bm{P}$ are zero, the probability of any measurement value falling in (belonging to) the corresponding domain (class) is zero. If the columns of $\bm{P}$ are linearly dependent, the probability of a sample belonging to class $C_k$ having a measurement in domain $D_j$ is a linear combination of the probabilities of samples belonging to all other classes having measurements in domain $D_j$. 
This occurs for a choice of partition where all points fall in a single domain $D_j$. In this extreme case, there is an apparent dependence (in the linear algebra sense) of the measurement values of different classes.  
As a related example, for the 1D SARS-CoV-2 antibody data from \cite{ainsworth2020performance} and \cite{wei2021antibody}, we can construct a partition where one trial domain is empty, both $\bm{P}$ and $\bm{P} - \bm{P_n}$ are singular, and therefore prevalence estimation is not possible. 

To avoid this situation, one should select nonempty trial domains, i.e., training data should lie in each element of the partition. 
In the limiting case that $P_{ij} = 0$, the measurement of a sample in class $C_j$ has zero probability of falling in domain $D_i$. The most extreme separation of training data occurs when the PDFs have nonzero support only on mutually exclusive elements of the partition. In this setting, the matrix $\bm{P}$ is a permutation matrix, and the prevalence estimates are merely the relative fractions $\bm{\hat{Q}}$ of measurements in each domain. If the partition elements are correctly matched to the classes, this extreme separation corresponds to a perfect assay because there are no misclassifications. 
We note that the matrices that result from a $k$-means partition of the 1D SARS-CoV-2 tri-class data are close to permutation matrices; one example is
\begin{equation}
\bm{P} = \left[\begin{array}{ccc}
0.9749 & 0.0058 & 0.1055 \\
0.0237 & 0.0495 & 0.8101 \\
0.0015 & 0.9447 & 0.0844
\end{array} \right].
\label{eq:ex_P}
\end{equation}
Selection of trial domains with a high degree of class separation may be a key to our low-error prevalence estimates.

We speculate that under certain conditions a matrix $\bm{P}$ that is a permutation matrix may be optimal in the sense that it minimizes the prevalence estimate error. 
In 1D, it may be possible to construct an optimization in terms of the samples assigned to each element of the partition, such as
\begin{equation}
\argmin_{f_1(x), f_2(x), \bm{P_{\pi}}} ||\bm{P_{\pi} P} - \bm{I} ||_2^2,
\end{equation} 
where $f_1(x)$ and $f_2(x)$ are indicator functions determining which samples are assigned to elements 1 and 2 of the partition (without loss of generality). 
Here, $\bm{P_{\pi} P}$ is the row permutation of $\bm{P}$ closest to the identity matrix, $\bm{I}$. For the matrix $\bm{P}$ given by (\ref{eq:ex_P}), for example, $\bm{P_{\pi}} = [\bm{e}_1; \bm{e}_3; \bm{e}_2]$, where $\bm{e}_j$ is the $j$th standard basis vector. 

We leave a search for the minimum prevalence error estimate to future work; see \cite{patrone2022minimizing} for an approach to the binary case. The extension of their work to the multiclass setting is not obvious because the objective function to minimize can be generalized in many different ways.

As a final note on extreme cases of prevalence estimation, in expectation, the problem is unconstrained. The constrained problem may be a viable alternative when it is known that one prevalence is close to zero.

\subsection{Local accuracy}
\label{sec:loc_acc}

Recall that in Section \ref{sec:opt_class} we needed to consider sets of measurements with equal probability of belonging to more than one class. This concept is related to \textit{local accuracy}, $Z$, which compares the probability that a test sample belongs to a particular class and has measurement $\bm{r}$ to the measurement density of a test sample with measurement $\bm{r}$.
We generalize the binary version from \cite{patrone2022holdout} to the multiclass setting:
\begin{equation}
Z(\bm{r}, D_1, \ldots, D_n) = \frac{q_k P_k(\bm{r})}{Q(\bm{r})} = \frac{q_k P_k(\bm{r})}{\sum_{j = 1}^n q_j P_j(\bm{r})}, \quad \bm{r} \in D_k,
\end{equation}
where $\{D_k\}$ partitions $\Omega$. Let $Z^{\star}(\bm{r}) = Z(\bm{r}, D_1^{\star}, \ldots, D_n^{\star})$ be the local accuracy of the optimal solution to the multiclass problem.  It is straightforward to show that $1/n \leq Z^{\star} \leq 1$. Due to optimality, if $\bm{r} \in D_k^{\star}$, we have $ q_k P_k(\bm{r}) \geq q_j P_j(\bm{r})$ for $j \neq k$.
Then 
\begin{equation}
n q_k P_k (\bm{r})  \geq \sum_{j = 1}^n q_j P_j(\bm{r}) = Q(\bm{r}),
\end{equation}
and so $q_k P_k(\bm{r})/Q(\bm{r}) =  Z^{\star}(\bm{r}) \geq 1/n$ for $r \in D_k^{\star}$. $Z^{\star}$ is maximized at 1 when $Q(\bm{r}) = q_k P_k(\bm{r})$ for $\bm{r} \in D_k^{\star}$. 
 In the multiclass setting, we have $Z^{\star} = 1/n$ when 
\begin{equation}
q_1 P_1(\bm{r}) = \ldots = q_n P_n(\bm{r}).
\end{equation}
We will refer to such an $\bm{r}$, if it exists, as a multipoint of the optimal domains. The lower bound on $Z^{\star}$ is only attained at a multipoint. To see this, consider some measurement $\bm{v}$ that is not a multipoint.  Then there exist $j, k \in \{1, \ldots, n\}$, $j \neq k$, such that $q_j P_j(\bm{v}) < q_k P_k(\bm{v})$. Then, since the classification is optimal, $\bm{v} \not \in D_j^{\star}$ and $\bm{v} \in D_m^{\star}$ for some $m$ (it may be that $m = k$). Clearly, $q_j P_j(\bm{v}) < q_m P_m(\bm{v})$. Further, $q_{\ell} P_{\ell} (\bm{v}) \leq q_m P_m(\bm{v})$ for $\ell \neq j$. It follows that
\begin{equation}
Q(\bm{v}) = \left[ \sum_{\substack{i = 1 \\ i \neq j}}^n q_i P_i(\bm{v}) \right] +  q_j P_j(\bm{v}) < (n-1) q_m P_m(\bm{v}) + q_m P_m (\bm{v})
\end{equation}
and so $Q(\bm{v}) < n q_m P_m(\bm{v})$, which gives $1/n < Z^{\star}(\bm{v})$ for a non-multipoint.

 The concept of local accuracy could be used to  decide which values to hold out in an indeterminate class in order to meet a global accuracy target \citep[see][]{patrone2022holdout}. 
For any measurement, we can compute what the local accuracy would be if we chose to assign it to each class in turn. Using the SARS-CoV-2 tri-class example as an illustration, conducting this procedure on a measurement $\bm{r}$ may result in, say, similarly high local accuracies for the previously infected and vaccinated classes, but a low local accuracy for the na{\"i}ve class. In this situation and without an optimal classification scheme, we may not feel confident labeling the sample as previously infected or as vaccinated, since their probabilities are similar, but we can say the sample is almost certainly not na{\"i}ve. This leads naturally to the observation that any subset of classes can be combined to make a new class. In particular, we can reduce the problem to a binary classifier. For our serological example, perhaps it is desirable to consider previously infected and vaccinated samples together, or equally possible that it is difficult to tell them apart for a particular assay, and so our goal becomes to classify them separately from na{\"i}ves. This reduction of the problem size by combining classes is in a sense a projection onto a lower class space. Specifically, consider
\begin{equation}
Q(\bm{r}) = \sum_{j = 1}^n q_j P_j(\bm{r}) = \underbrace{\sum_{j = 1}^k q_j P_j (\bm{r})}_{\tilde{q}_1 \tilde{P}_1(\bm{r})} + \underbrace{\sum_{j = k+1}^n q_j P_j (\bm{r})}_{\tilde{q}_2 \tilde{P}_2(\bm{r})},
\end{equation}
where $\tilde{P}_1(\bm{r})$ and $ \tilde{P}_2(\bm{r})$ are newly-created PDFs with associated prevalences $\tilde{q}_1$ and $\tilde{q}_2 = 1 - \tilde{q}_1$. The task becomes to find $\tilde{q}_1$, for which there may be an optimal strategy. 

The diagnostic community's current analog to local accuracy is the concept of a likelihood ratio (LR), calculated as $s_e/(1 - s_p)$ for a previously infected test result, where $s_e$ and $s_p$ represent sensitivity and specificity. The previously infected LR can be interpreted as the ratio of probabilities of correctly to incorrectly predicting a previously infected result \citep{riffenburgh2011statistics}. These values use average population information through $s_e$ and $s_p$ values, which may not always be available or representative. In contrast, local accuracy uses local information, since the latter is conditioned on knowing individual measurement values.

\subsection{Extensions}

Multiclass methods are readily equipped to handle further stratification of antibody data, such as by age group, biological sex, or coronavirus disease of 2019 (COVID-19) booster status. An additional class could be added for individuals who are both vaccinated and previously infected. Studies have demonstrated a greater antibody response post-vaccination for previously-infected versus COVID-19 na\"{i}ve recipients \citep{dalle2021serology,narowski2022sars}; this could allow for these populations to be distinguished by our classification scheme. Further, we minimize the prevalence-weighted combination of misclassifications, but the optimization problem can be rewritten for any desired objective function. Reformulations include ``rule-in'' or ``rule-out'' tests that meet desired sensitivity or specificity targets \citep{florkowski2008sensitivity}. Our methods may even be generalizable to multi-label classification, in which a sample can be assigned to more than one class; we anticipate challenges designing the corresponding optimization problem. Finally, the methods presented here can be applied to any setting where class size estimation and population labeling are required; an example is cell sorting in flow cytometry. 

\subsection{Limitations}

Model selection is inherently subjective;  \cite{schwartz1967estimation} showed that the error goes to zero as more data points are added. As the number of antibody measurements increases, corresponding to viewing the data in higher dimensions, additional modeling choices become available.
 \cite{patrone2021classification} suggest the possibility of minimizing misclassifications over a family of models; see also \cite{smith2013uncertainty} for a discussion of model form errors. Classification accuracy and prevalence estimation of the 1D data sets from \cite{ainsworth2020performance} and \cite{wei2021antibody} suffer from overlap between their spike IgG values. If more measurements were available per sample, modeling the data in a higher dimension could improve class separation and thereby lower error rates \citep[see][]{luke2022improving}. Further, our models do not account for time-dependence. This concept is important when classifying antibody tests, which are known to have a half life on the order of several months post infection or vaccination \citep{xia2021longitudinal,kwok2022waning}. See \cite{bedekar2022prevalence} for a time-dependent approach to the binary setting.

\subsection{Implications for assay developers}

We have solved the multiclass diagnostic classification problem, which was previously unresolved. Antibody measurements from vaccinated individuals can now be distinguished from previously infected and na{\"i}ve samples.

Our work is the first to obtain unbiased predictions of the relative fractions of vaccinated, previously infected, and na{\"i}ve individuals in a population. These estimates are improved as more samples are added.  Best practices for conducting these predictions include dividing the range of all possible measurement values into nonempty regions that create separation between samples of neighboring regions. This can be easily achieved using pre-defined clustering algorithms. Our procedure hinges on selecting probability distributions to model training populations, which can be conducted automatically for measurements of a single antibody target in several open-source programming languages. Our classification scheme is also easily implementable, and can be modified to prioritize specificity if desired. Regardless of the reformulation, the error is minimized by construction. 

\section{Acknowledgements}

This work is a contribution of the National Institute of Standards and Technology and is not subject to copyright in the United States. R.L. was funded through the NIST PREP grant 70NANB18H162. The aforementioned funder had no role in study design, data analysis, decision to publish, or preparation of the manuscript. Use of data provided in this paper has been
approved by the NIST Research Protections Office (IRB no. ITL 2020 2057). The authors wish to thank Drs. Daniel Anderson and Eric Shirley for useful discussions during preparation of this manuscript.

\section{Data Availability}

Analysis scripts and data developed as a part of this work are available upon reasonable request. Original data are provided in \cite{ainsworth2020performance} and \cite{wei2021antibody}.

\section{Declarations of Competing Interests}

The authors have no competing interests to declare.

\setcounter{equation}{0}

\appendix

\renewcommand{\theequation}{A\arabic{equation}}

\section{Optimality of classification domains}
\label{sec:app_a}

\begin{lem}
Let the PDFs $P_j(\bm{r})$ be bounded and summable functions on $\Omega$ and suppose that the measure of a single measurement $\bm{r}$ is zero with respect to all distributions. Further, suppose the boundary set $\mathscr{E}_j$ has measure zero for all $j$. Define the loss function $\mathscr{L}$ as in (\ref{eq:loss}). Then the domains given by (\ref{eq:opt_d})
minimize the loss function $\mathscr{L}$.
\end{lem}

\begin{proof}
Consider a partition of $\Omega$ given by $\{D_j^{\star}\}$ for which each $D_j^{\star}$ satisfies the criteria given in Eqs. (\ref{eq:req}) and (\ref{eq:opt_d}). 
Without loss of generality, absorb each $q_j$ into $P_j(\bm{r})$, and consider the loss function
\begin{equation}
\mathscr{L} (D_1, \ldots, D_n) = \sum_{j = 1}^n \int_{\Omega \setminus D_j} P_j (\bm{r}) d\bm{r}.
\label{eq:loss_func}
\end{equation}
Now consider a different partition of the measurement space $\{\hat{D}_j\}$, where at least two elements $\hat{D}_k$ and $\hat{D}_i$ differ from $D_k^{\star}$ and $D_i^{\star}$ 
by more than a set of measure zero\footnote{If only one element $\hat{D}_k$ differed by more than a set of measure zero from $\hat{D}_k^{\star}$, the set $\{\hat{D}_j\}$ would no longer form a partition of $\Omega$.}. Without loss of generality, suppose the the elements of the partitions $\{\hat{D}_j\}$ and $\{D_j^{\star}\}$ are ordered such that $\hat{D}_1, \ldots \hat{D}_M$ differ from $D_1^{\star}, \ldots, D_M^{\star}$ by more than a set of measure zero and $\hat{D}_{M+1}, \ldots \hat{D}_n$ and $D_{M+1}^{\star}, \ldots, D_n^{\star}$ do not. 
Note that $\Omega \setminus \hat{D}_j  = (\hat{D_j})^C$.
 As a slight abuse of notation, we will rewrite $(\hat{D_j})^C$ and subsequent, similar expressions as
\begin{equation}
 (\hat{D_j})^C = \bigcup_{\substack{k = 1 \\ k \neq j}}^n \hat{D_k}
\end{equation}
with the understanding that this is true up to sets of measure zero and that the discrepancy does not affect integration.
Then we may decompose the loss function (\ref{eq:loss_func}) as
\begin{equation}
\mathscr{L}(\hat{D_1}, \ldots, \hat{D_n}) = \sum_{j =1}^n \int_{\cup_{k, k \neq j} \hat{D}_{k}} P_j(\bm{r}) d\bm{r} .
\label{eq:L_hat}
\end{equation}
Likewise,
\begin{equation}
\mathscr{L}(D_1^{\star}, \ldots, D_n^{\star}) = \sum_{j =1}^n \int_{\cup_{k, k \neq j} {D}^{\star}_{k}} P_j(\bm{r}) d\bm{r}.
\label{eq:L_star}
\end{equation}
 Decomposing $(\hat{D}_j)^C$ further gives
\begin{equation}
(\hat{D}_j)^C = \bigcup_{\substack{k = 1 \\ k \neq j}}^n \hat{D_k} = \bigcup_{\substack{k = 1 \\ k \neq j}}^n \left[ (\hat{D_k} \cap  D_k^{\star} ) \cup (\hat{D_k} \setminus D_k^{\star}) \right]  = \left[ \bigcup_{\substack{k = 1 \\ k \neq j}}^n (\hat{D_k} \cap D_k^{\star}) \right] \cup \left[ \bigcup_{\substack{k = 1 \\ k \neq j}}^n (\hat{D_k} \setminus D_k^{\star})\right].
\label{eq:decomp}
\end{equation}
Here, (\ref{eq:decomp}) can be further decomposed using our assumptions on which elements of the partitions are equivalent up to sets of measure zero:
\begin{equation}
\begin{split}
(\hat{D}_j)^C  & = \left[ \bigcup_{\substack{k = 1 \\ k \neq j}}^M (\hat{D_k} \cap D_k^{\star}) \right] \cup \left[ \bigcup_{\substack{k = 1 \\ k \neq j}}^M (\hat{D_k} \setminus D_k^{\star})\right]  \cup \left[ \bigcup_{\substack{k = M+1 \\ k \neq j}}^n (\hat{D_k} \cap D_k^{\star}) \right] \cup \left[ \bigcup_{\substack{k = M+1 \\ k \neq j}}^n (\hat{D_k} \setminus D_k^{\star}) \right],
\end{split}
\label{eq:decomp1}
\end{equation}
 The third term in square brackets in (\ref{eq:decomp1}) has the same measure as $\cup_{k = M+1, k \neq j}^n \hat{D}_k$ and the last term has measure zero by our assumptions. We let the first term
 be denoted by $\hat{D}_{j \cap}$, the second by $\hat{D}_{j \setminus}$, and the third by $\hat{D}_{j \cup}.$ Define $D_{j \cap}^{\star}$, $D_{j \setminus}^{\star}$, and $D_{j \cup}^{\star}$ similarly as the first, second, and third terms in (\ref{eq:decomp1}) but with $\wedge$ and $\star$ swapped. Note that
 all terms in (\ref{eq:decomp1}) are disjoint. 
Thus, subtracting (\ref{eq:L_star}) from (\ref{eq:L_hat}) gives
\begin{equation}
\begin{split}
\Delta \mathscr{L} & = \mathscr{L}(\hat{D_1}, \ldots, \hat{D_n}) - \mathscr{L}(D_1^{\star}, \ldots, D_n^{\star}) \\
& =  \sum_{j =1}^n \int_{\hat{D}_{j \cap}} P_j(\bm{r}) d\bm{r} + \sum_{j =1}^n \int_{\hat{D}_{j \setminus}} P_j(\bm{r}) d\bm{r} + \sum_{j = 1}^n \int_{\hat{D}_{j \cup}} P_j(\bm{r}) d\bm{r} \\
&  - \sum_{j =1}^n \int_{D_{j \cap}^{\star}} P_j(\bm{r}) d\bm{r} -  \sum_{j =1}^n \int_{D_{j \setminus}^{\star}} P_j(\bm{r}) d\bm{r} - \sum_{j = 1}^n \int_{D_{j \cup}^{\star}} P_j(\bm{r}) d\bm{r} \\
& = \sum_{j = 1}^n \int_{\hat{D}_{j \setminus}} P_j(\bm{r}) d \bm{r} - \sum_{j = 1}^n \int_{D_{j \setminus}^{\star} } P_j(\bm{r}) d \bm{r},
\end{split}
\end{equation}
where we have used our assumption that $
\hat{D}_{j \cup} =  D_{j \cup}^{\star}$ up to sets of measure zero and
\begin{equation}
\hat{D}_{j \cap} = \bigcup_{\substack{k = 1 \\ k \neq j}} (\hat{D_k} \cap D_k^{\star}) = \bigcup_{\substack{k = 1 \\ k \neq j}} ( D_k^{\star} \cap \hat{D_k}) = D_{j \cap}^{\star}.
\end{equation} 
Since$\{ \hat{D_j} \}$ and $\{D_j^{\star}\}$ partition $\Omega$, we may write
\begin{equation}
\Delta \mathscr{L} = \sum_{j = 1}^n \sum_{\substack{k = 1 \\ k \neq j}}^M \left[ \int_{\hat{D_k} \setminus D_k^{\star}} P_j(\bm{r}) d \bm{r} - \int_{D_k^{\star} \setminus \hat{D_k}} P_j(\bm{r}) d \bm{r} \right].
\label{eq:DL_Dk}
\end{equation}
From our assumption that $\hat{D_j}$ differs from $D_j^{\star}$ by a measurable set for all $j \in \{1, \ldots, M\}$, each set $\hat{D_k} \setminus D_k^{\star}$ in $\hat{D}_{j \setminus}$ and $D_i^{\star} \setminus \hat{D_i}$ in $D_{j \setminus}^{\star}$ have nonzero measure.  Next, we may write
\begin{equation}
\begin{split}
\hat{D}_{j \setminus} = \bigcup_{\substack{k = 1 \\ k \neq j}}^M \hat{D_k} \setminus D_k^{\star} = \bigcup_{\substack{k = 1 \\ k \neq j}}^M \left\{ \left[ ( \hat{D_k} \setminus D_k^{\star}) \cap D_j^{\star} \right] \cup \left[ ( \hat{D_k} \setminus D_k^{\star} ) \setminus D_j^{\star}\right] \right\}  \\= \left\{\bigcup_{\substack{k = 1 \\ k \neq j}}^M \left[ ( \hat{D_k} \setminus D_k^{\star}) \cap D_j^{\star} \right] \right\} \cup \left\{\bigcup_{\substack{k = 1 \\ k \neq j}}^M \left[ ( \hat{D_k} \setminus D_k^{\star} ) \setminus D_j^{\star}\right] \right\} , 
\end{split}
\label{eq:split1}
\end{equation}
and similarly for $ D_{j \setminus}^{\star} =  \bigcup_{\substack{k = 1 \\ k \neq j}} D_k^{\star} \setminus \hat{D_k}$. Call the first term in curly braces in (\ref{eq:split1}) $\hat{D}_{ j \setminus \cap}$ and the second $\hat{D}_{j \setminus \setminus}$, and let $D_{j \setminus \cap}^{\star}$ and $D_{j \setminus \setminus}^{\star} $ be similarly defined. Note that $\hat{D}_{j \cap \setminus}$ and $D_{j \setminus \setminus}^{\star}$ are disjoint by construction. This allows us to write
\begin{equation}
\Delta \mathscr{L} = \sum_{j = 1}^n \sum_{\substack{k = 1 \\ k \neq j}}^M \left[ \int_{\hat{D}_{ j\setminus \cap}} P_j(\bm{r}) d \bm{r} + \int_{\hat{D}_{j \setminus \setminus}} P_j(\bm{r}) d \bm{r}  - \int_{D_{j \setminus \cap}^{\star}} P_j(\bm{r}) d \bm{r} - \int_{D_{j \setminus \setminus}^{\star}} P_j(\bm{r}) d \bm{r} \right].
\label{eq:DL_Dk_Di}
\end{equation}
We rewrite $(D_j^{\star})^C$ to find
\begin{equation}
(\hat{D_k} \setminus D_k^{\star}) \setminus D_j^{\star} = (\hat{D_k} \setminus D_k^{\star}) \cap \left[ \bigcup_{\substack{i = 1 \\ i \neq j}}^n D_i^{\star} \right]  = \hat{D_k} \cap \left[ \bigcup_{\substack{\ell = 1 \\ \ell \neq k}}^n D_{\ell}^{\star} \right] \cap \left[ \bigcup_{\substack{i = 1 \\ i \neq j}}^n D_i^{\star} \right]  = \bigcup_{\substack{i = 1 \\ i \neq j, k}}^n \hat{D_k} \cap D_{i}^{\star}.
\label{eq:DJC}
\end{equation}
We rewrite $(\hat{D}_j)^C$ similarly. Here, (\ref{eq:DJC}) and the symmetry of the intersection operator give
\begin{equation}
\hat{D}_{j \setminus \setminus}  = \bigcup_{\substack{i = 1 \\ i \neq j, k}}^n  \hat{D_k} \cap D_i^{\star}  =  \bigcup_{\substack{i = 1 \\ i \neq j, k}}^n   D_k^{\star} \cap \hat{D}_i = D_{j \setminus \setminus}^{\star}.
\end{equation}
 Thus, these terms in (\ref{eq:DL_Dk_Di}) are equal and of opposite signs, and cancel. This leaves
\begin{equation}
\Delta \mathscr{L} = \sum_{j = 1}^n \sum_{\substack{k = 1 \\ k \neq j}}^M \left[ \int_{ \hat{D}_{j \setminus \cap}} P_j(\bm{r}) d \bm{r} - \int_{ D_{j \setminus \cap}^{\star}} P_j(\bm{r}) d \bm{r} \right].
\label{eq:DL_2}
\end{equation}
Since
\begin{equation}
 D_{j \setminus \cap}^{\star} =  \bigcup_{\substack{k = 1 \\ k \neq j}} (D_k^{\star} \setminus \hat{D_k}) \cap \hat{D_j}  \subset (D_j^{\star})^C,
\end{equation}
we have $\mu( D_{j \setminus \cap}^{\star} \cap D_j^{\star} ) \leq \mu ((D_j^{\star})^C \cap D_j^{\star})  = 0 .$
 Thus, since there is no measurable overlap with the maximal set $D_j^{\star}$, we have $P_j(\bm{r}) \leq P_k(\bm{r})$ on $D_{j \setminus \cap}^{\star}$ for $k \neq j$. In contrast, $ \hat{D}_{j \setminus \cap} \cap D_j^{\star} = (\hat{D_k} \setminus D_k^{\star}) \cap D_j^{\star}$ is by construction a subset of $D_j^{\star}$, and in totality $ \bigcup_j \hat{D}_{j \setminus \cap} \cap D_j^{\star} \neq \emptyset$ since we have assumed that $\hat{D}_1, \ldots, \hat{D}_M$ differ from $D_1^{\star}, \ldots, D_M^{\star}$ by a set of nonzero measure. Thus $P_j(\bm{r}) \geq P_k(\bm{r})$ on $\hat{D}_{j \setminus \cap}$ for  $k \neq j$. 

Since $P_j$ is maximal on each $\hat{D}_{j \setminus \cap}$ and submaximal on each $D_{j \setminus \cap}^{\star}$, and $\bigcup_j \hat{D}_{j \setminus \cap}$ has nonzero measure, we have
\begin{equation}
\Delta \mathscr{L} = \sum_{j = 1}^n \sum_{\substack{k = 1 \\ k \neq j}}^M \left[ \int_{ \hat{D}_{j \setminus \cap}} P_j(\bm{r}) d \bm{r} - \int_{ D_{j \setminus \cap}^{\star}} P_j(\bm{r}) d \bm{r} \right] \geq 0,
\end{equation}
which proves that $\{D_j^{\star}\}$ as defined minimizes the loss function $\mathscr{L}$.
\end{proof}

\renewcommand{\theequation}{B\arabic{equation}}

\section{Measurement details}
\label{sec:app_MFI}

\cite{ainsworth2020performance} provide previously infected and na{\"i}ve samples (which they refer to as positive and negative). 
\cite{ainsworth2020performance} used a true negative rate of 99 \% to set a positivity threshold of 8 million MFI units. Previously infected serological samples were taken at least 20 days post symptom onset from individuals whose infections were confirmed via reverse transcription polymerase chain reaction (RT-PCR) by a nose or throat swab. The na{\"i}ve serological samples were collected pre-pandemic. 

\cite{wei2021antibody} provide vaccinated serological samples. Measurements were recorded from individuals that had been innoculated with either the Oxford--AstraZeneca ChAdOx1 nCoV-19 (Vaccine A) or the Pfizer-BioNTech BNT162b2 (Vaccine B). The vaccinated samples were collected at various time points relative to  the first inoculation, ranging from 14 days prior to 66 days after. To control for variation resulting from the length of time after inoculation, we use only data from 28 days post first dose. 

 \cite{ainsworth2020performance} recorded antibody measurements in MFI. For comparison, \cite{wei2021antibody} convert from MFI to ng/mL following:
\begin{equation}
\log_{10}(m) = A + B f + C \mathbb{I}_{f > D} (f - D),
\end{equation}
where the constants $A, B, C$, and $D$ are given by $A =0.221738, B =1.751889 \times 10^{-7}, C =   5.416675 \times 10^{-7}$, and $D =9.19031 \times 10^{6}$. Here, $m$ is the measurement in ng/mL, $f$ is the measurement in MFI, and $\mathbb{I}$ denotes the indicator function on all MFI values, returning one if the measurement is above $D$ and zero otherwise. \cite{wei2021antibody} truncated vaccinated measurements below $2$ ng/mL (0.4 \% of total data) and above $500$ ng/mL (7 \% of total data). The truncation was not applied to the previously infected and na{\"i}ve samples because they were first reported in \cite{ainsworth2020performance}. 

In total, there are 976 na{\"i}ve, 536 previously infected, and 686 vaccinated samples (362 Vaccine A and 324 Vaccine B samples taken at the 28 day mark). 

\renewcommand{\theequation}{C\arabic{equation}} 

\section{Data censoring}
\label{sec:app_trunc}

Given a general distribution $f(x; \bm{\varphi})$ of parameters $\bm{\varphi}$ that fits the data between the truncation limits $x_{\min} < x < x_{\max}$, the probability that a measurement is above the upper truncation limit is
\begin{equation} p_{\max} = \int_{x_{\max}}^{\infty} f(x) dx,
\end{equation}
and the probability that a measurement is below the lower truncation limit is
\begin{equation}
 p_{\min} = \int_{-\infty}^{x_{\min}} f(x) dx.
 \end{equation}
 These definitions of $p_{\max}$ and $p_{\min}$ are used in the updated likelihood function
\begin{equation}
L(x) = \begin{cases}
p_{\min}, & x = x_{\min}, \\
f(x), & x_{\min} < x < x_{\max}, \\
p_{\max}, & x = x_{\max}
\end{cases}.
\end{equation}
The PDF can be written as 
\begin{equation}
\hat{f}(x) = f(x) \mathbb{I}(x \in (x_{\min}, x_{\max})) + \delta(x - x_{\min}) p_{\min} + \delta(x - x_{\max}) p_{\max}.
\end{equation}

\renewcommand{\theequation}{D\arabic{equation}}

\section{2D synthetic probability models}
\label{sec:app_b}

Our 2D exploration builds off work by \cite{patrone2021classification}, which used data from an assay developed in \cite{liu2020quantification}. The two dimensions correspond to measurements taken at the receptor-binding domain (RBD), a substructure of the spike protein of the SARS-CoV-2 molecule, and the nucleocapsid (N) that stabilizes the RNA; values are recorded as MFIs. 
Values are recorded as MFIs but follow the same logarithmic scale given by (\ref{eq:log_transform}). The log-transformed RBD values are $x$ and N values are $y$.
All model parameters are determined via MLE. The na{\"i}ve samples should have relatively low values of both RBD and N since they have no specific immune response to SARS-CoV-2. However, we still expect small na{\"i}ve signals due to cross-reactivity with other coronaviruses such as NL63 and HKU1, which are versions of the common cold. In contrast, previously infected samples should have produced a relatively high immune response to both RBD and N.

It is natural to expect some correlation between the signals, which motivates a change of variables
$z = (x + y)/\sqrt{2},  w  = (x - y)/\sqrt{2}.$
This will cause the data to be distributed along the diagonal. Our na{\"i}ve distribution $N$ is given by
\begin{equation}
N(z, w; k, \alpha, \theta, \beta, \mu) = \frac{z^{k-1}}{\sqrt{2 \pi} \alpha \Gamma(k) \theta^k} \exp\l - \frac{z}{\theta} - \frac{z}{\beta} - \frac{(w - \mu)^2}{2 \alpha^2 \exp(2z/\beta)}\r.
\end{equation}
 This is a hybrid gamma-normal distribution where the variance of the variable corresponding to the direction perpendicular to the diagonal, $w$, depends on the variance of the variable corresponding to the diagonal, $z$:
$\sigma_w = \alpha \exp(z/\beta).$ 
This allows for the data to fan out slightly away from the origin along the diagonal.

The previously infected distribution $P$ is given by
\begin{equation}
P(\zeta, w; \alpha, \beta, \theta, \mu) = \frac{\Gamma(\alpha + \beta)}{\theta \sqrt{2 \pi \zeta}\Gamma(\alpha) \Gamma(\beta)} \zeta^{\alpha - 1} ( 1 - \zeta)^{\beta - 1} \exp \l - \frac{(w - \mu)^2}{2 \theta^2 \zeta}\r.
\end{equation}
This is a hybrid beta-normal distribution. Here, we use a different change of variables for the diagonal direction to rescale the beta distribution to its domain $[0, 1]$:
$\zeta =(x + y)/(9 \sqrt{2}).$
Again, the variance of the second variable $w$ depends on the first, $\zeta$:
$\sigma_w = \theta \sqrt{\zeta}.$
This is a modeling choice that allows for a slightly wider distribution for large N and RBD values. The beta distribution is selected because we recognize that the previously infected samples should span the entire diagonal line $y = x$.

When creating a synthetic vaccinated category, we consider characteristics that set vaccinated and previously infected individuals' antibody measurements apart. Since the vaccines target the binding site on the spike protein, we expect both vaccinated and previously infected individuals to have high RBD values.  However, only naturally infected individuals should exhibit high N values. Thus, we create a vaccinated category that is clustered in the bottom right of an N vs. RBD plot.

For the vaccinated ($V$) population, we use a hybrid Weibull-normal distribution: 
\begin{equation}
V(z, w; \alpha, \beta, \delta, \lambda, \mu) =  \frac{\delta}{\lambda \alpha \sqrt{2 \pi}} \l \frac{z}{\lambda} \r^{\delta - 1}  \exp \left[ - \left(\frac{z}{\lambda} \right)^{\delta}- \frac{z}{\beta} - \frac{(w-\mu)^2}{2 \alpha^2 \exp(2z/\beta) } \right].
\end{equation}
 The variance of the second variable $w$ depends on the first, $z$, and is the same form as that for the na{\"i}ve distribution.

 \bibliographystyle{spbasic}      % basic style, author-year citations

\bibliography{Multiclass_bib}   % name your BibTeX data base

\newpage
 
 \section{Supplemental data}
 
 \renewcommand{\thefigure}{S\arabic{figure}}

\begin{figure}[h]
\centering
\subfloat[][Vaccine B]{\includegraphics[scale=.5]{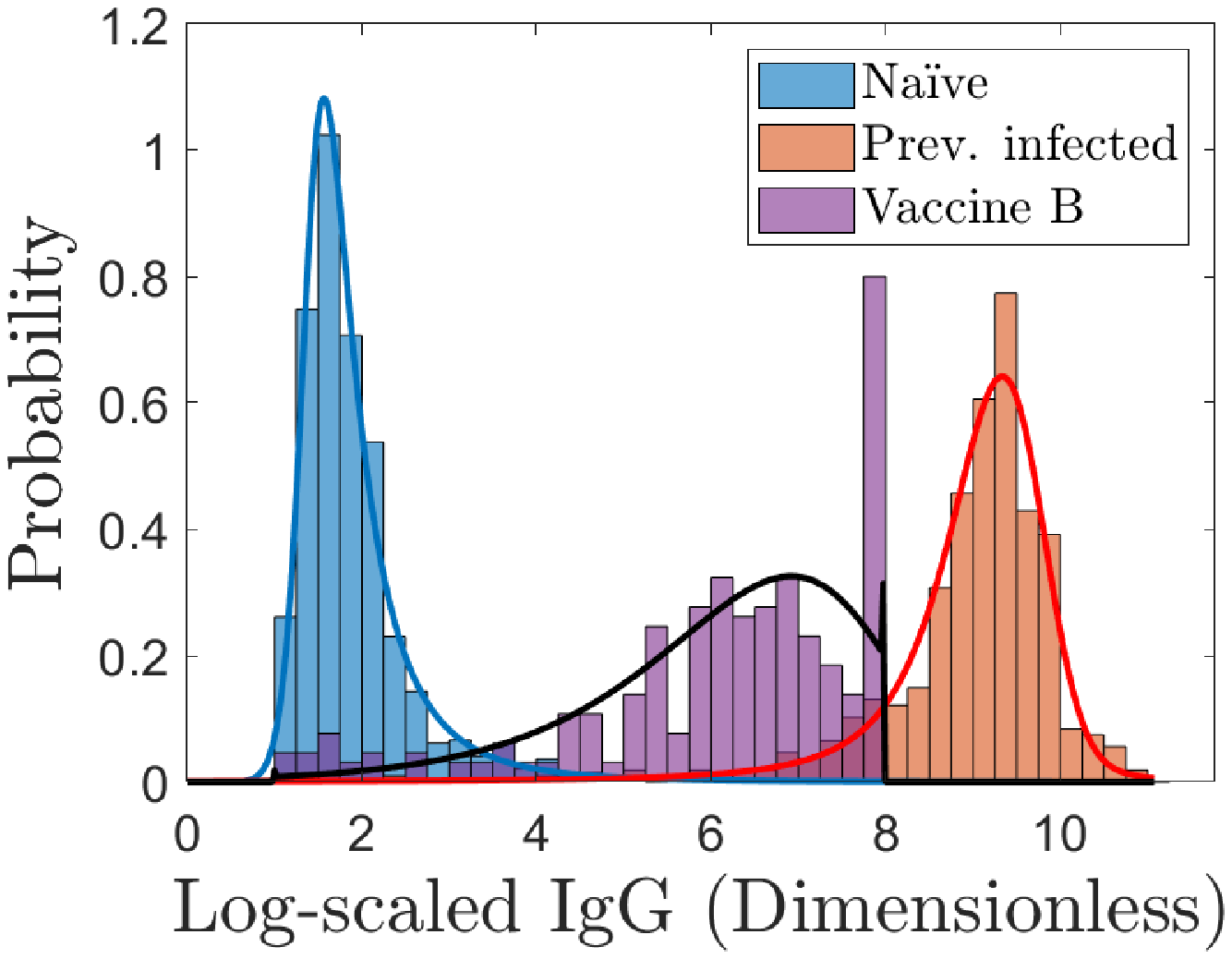}}
\subfloat[][Combined]{\includegraphics[scale=.5]{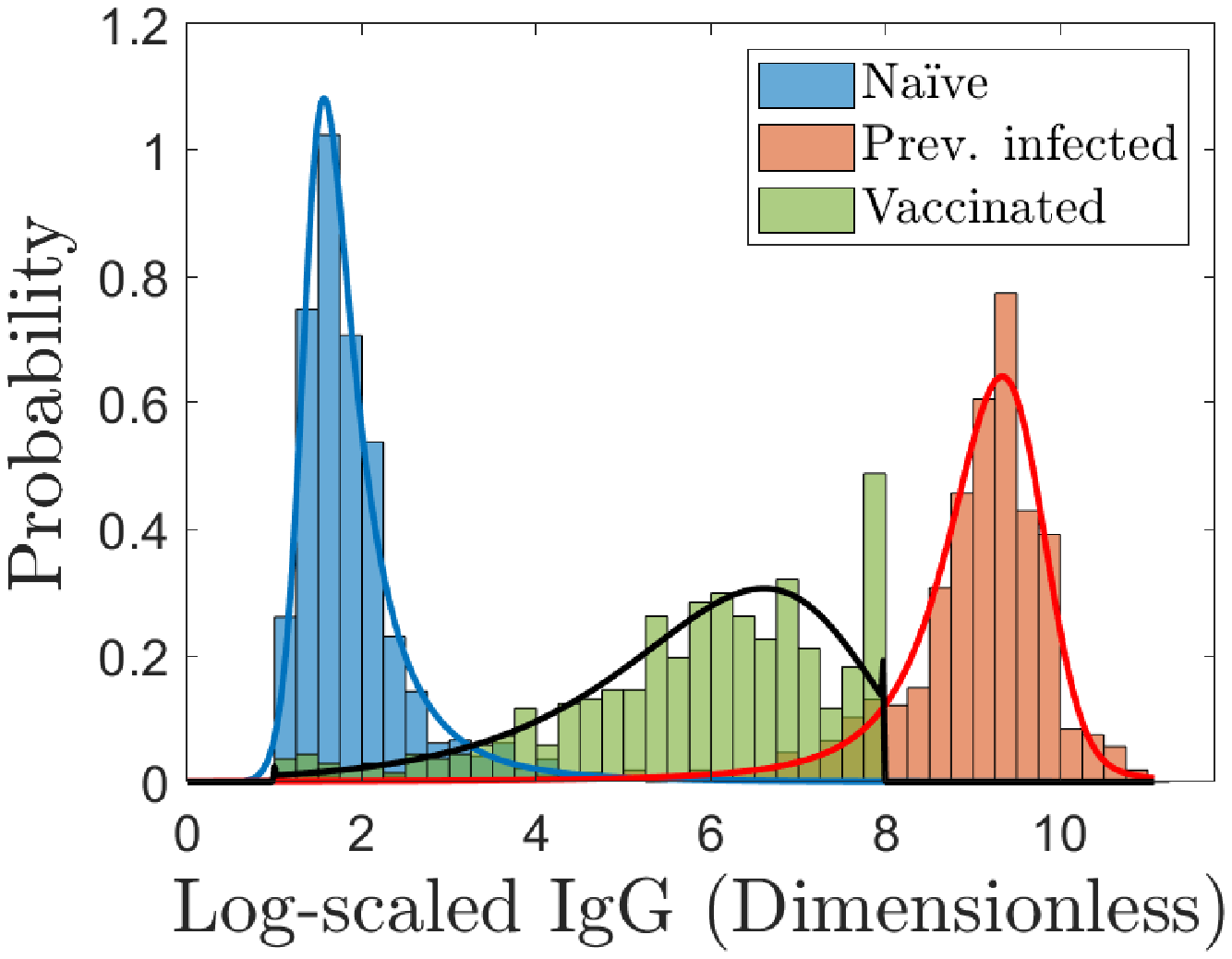}}
\caption{Conditional PDFs for the three classes trained on the training data for Vaccine B and combined visualizations of the vaccinated class.}
\label{fig:pdfs}
\end{figure}

\begin{figure}[H]
\centering
\subfloat[][Vaccine B]{\includegraphics[scale=.5]{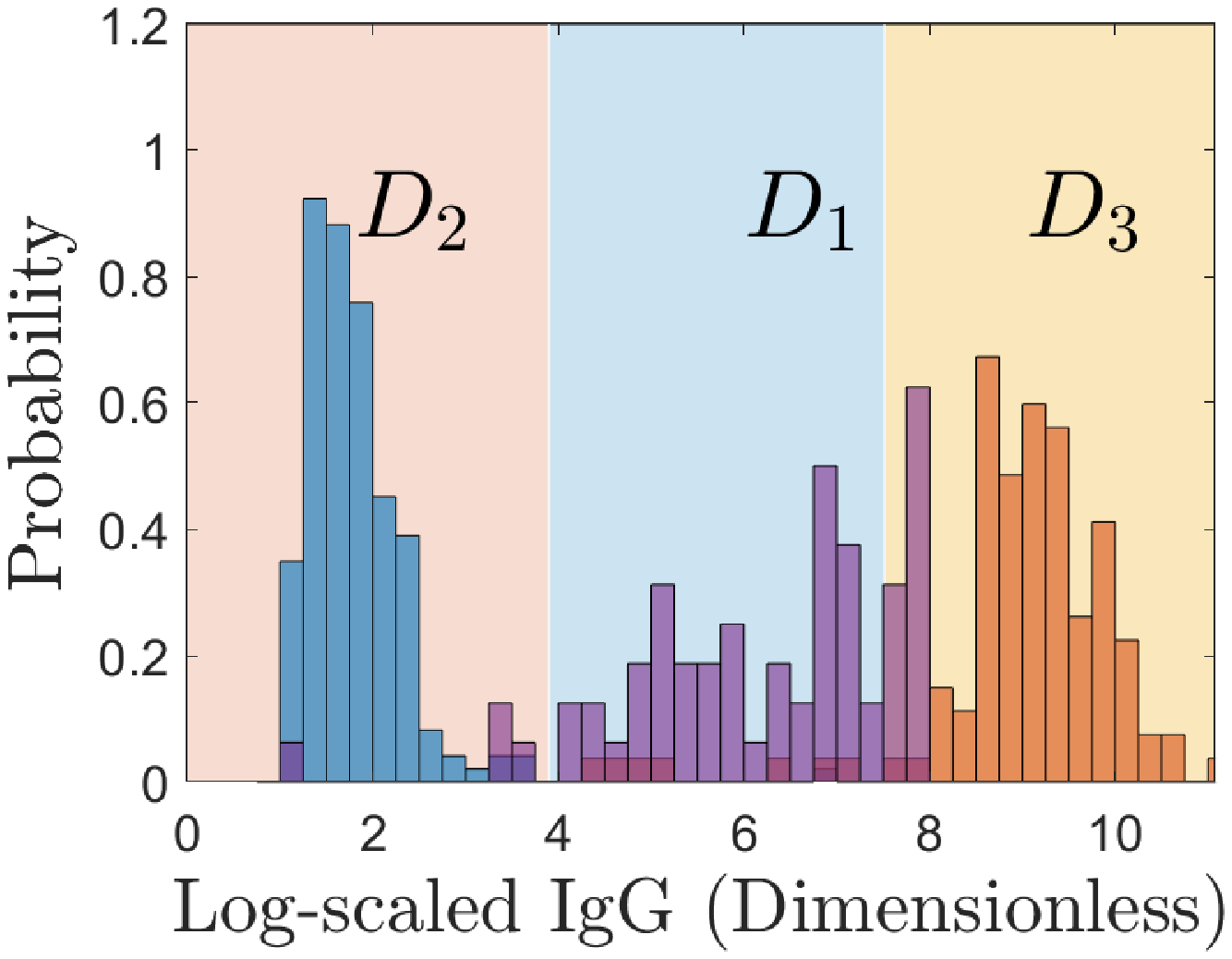}} 
\subfloat[][Combined]{\includegraphics[scale=.5]{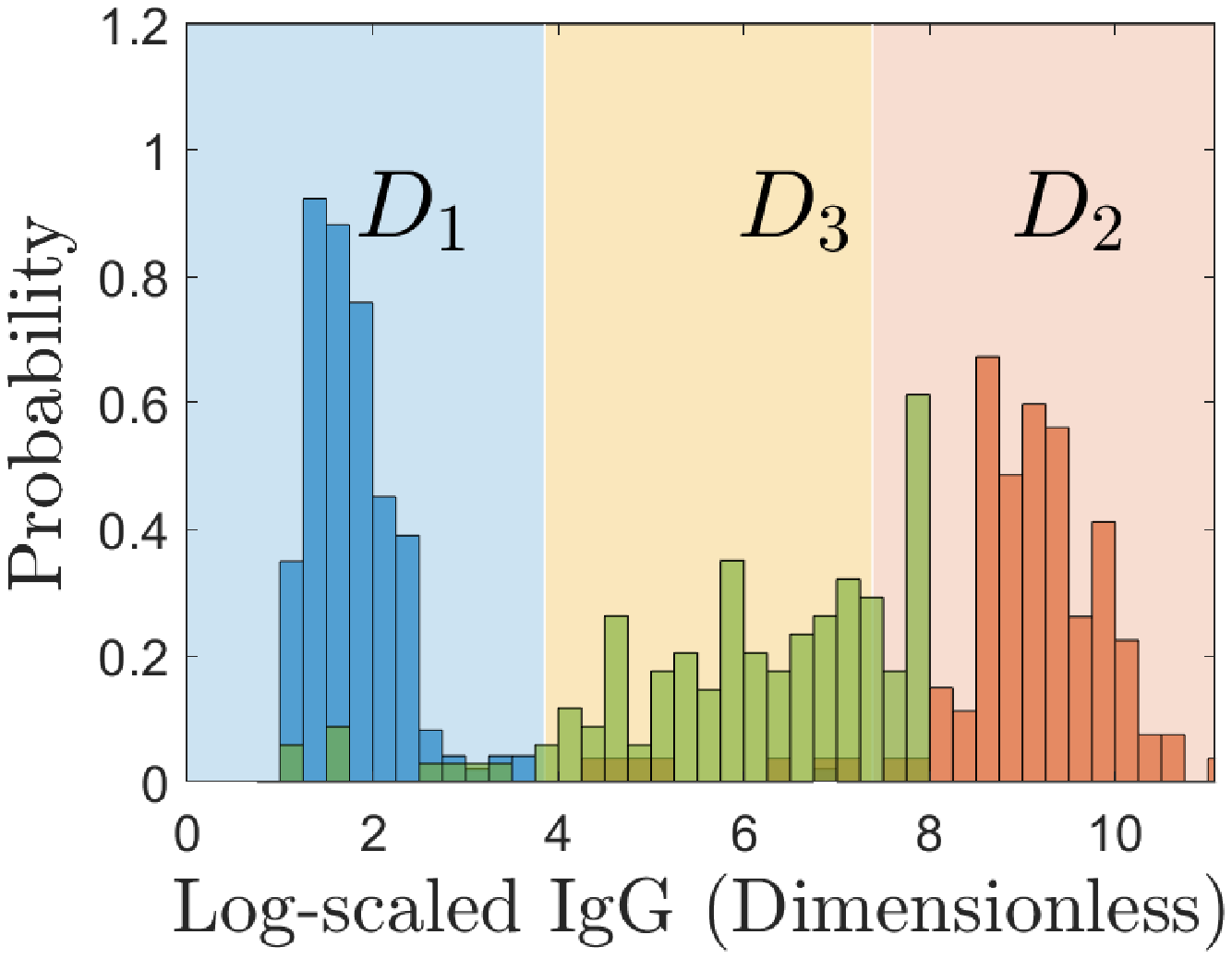}} 
\caption{Test data $k$-means partitioning for  generalized prevalence estimates. We use $k = 3$ classes; the clustered domains are labeled as $D_1, D_2$, and $D_3$.}
\label{fig:part_sup}
\end{figure}

Figure \ref{fig:pdfs} shows the conditional PDFs for the na{\"i}ve, previously infected, and vaccinated classes for two visualizations of the vaccinated class. The effects of truncating the data at the upper limit are visible in the right-most bins of the vaccinated class histograms; these are accounted for by the data-censored vaccinated PDFs. Figure \ref{fig:part_sup} shows the $k$-means partition for the Vaccine B and combined visualizations of the vaccinated class.

\begin{figure}[h]
\centering
\subfloat[][Training data]{\includegraphics[scale=.5]{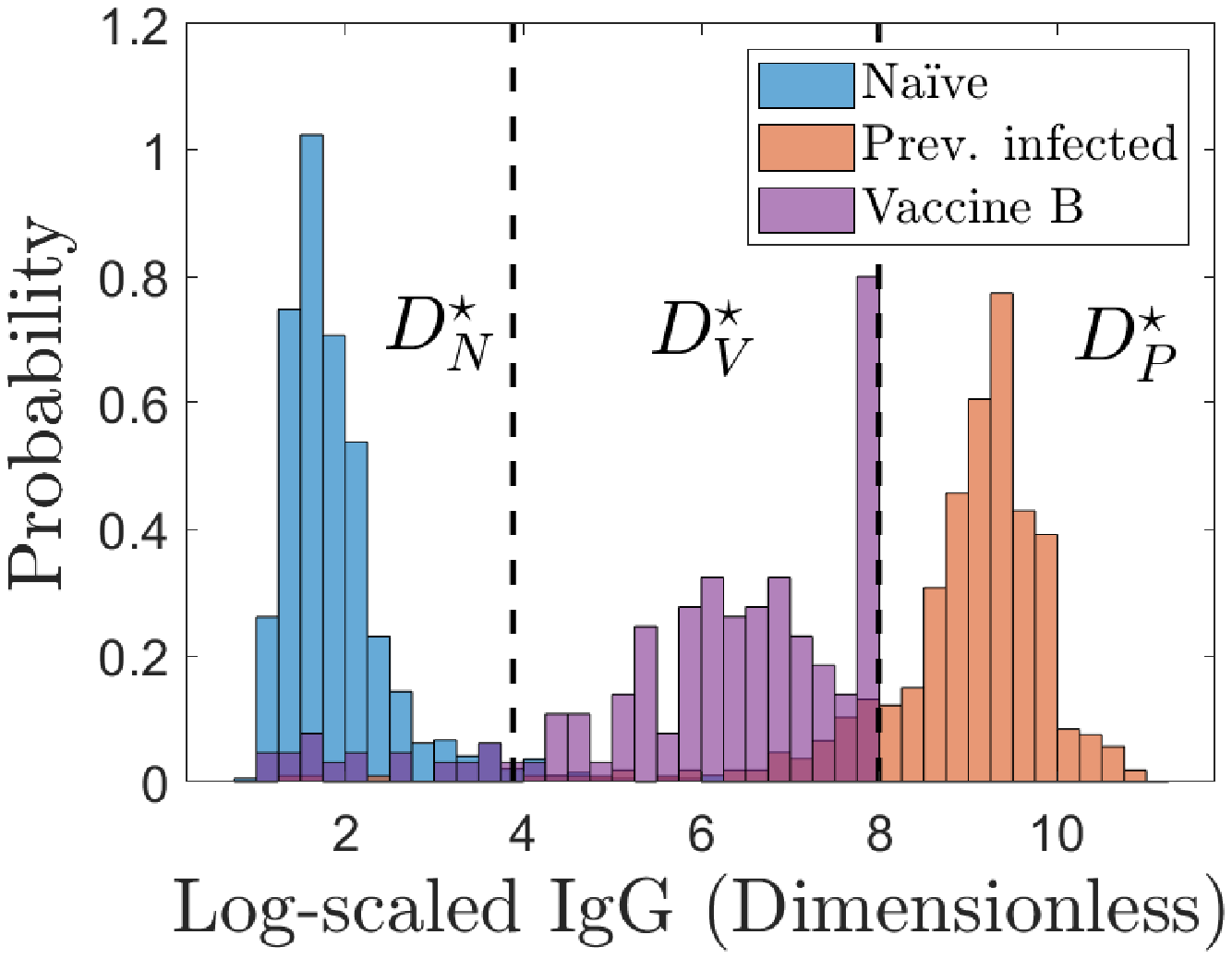}}  
\subfloat[][Test data]{\includegraphics[scale=.5]{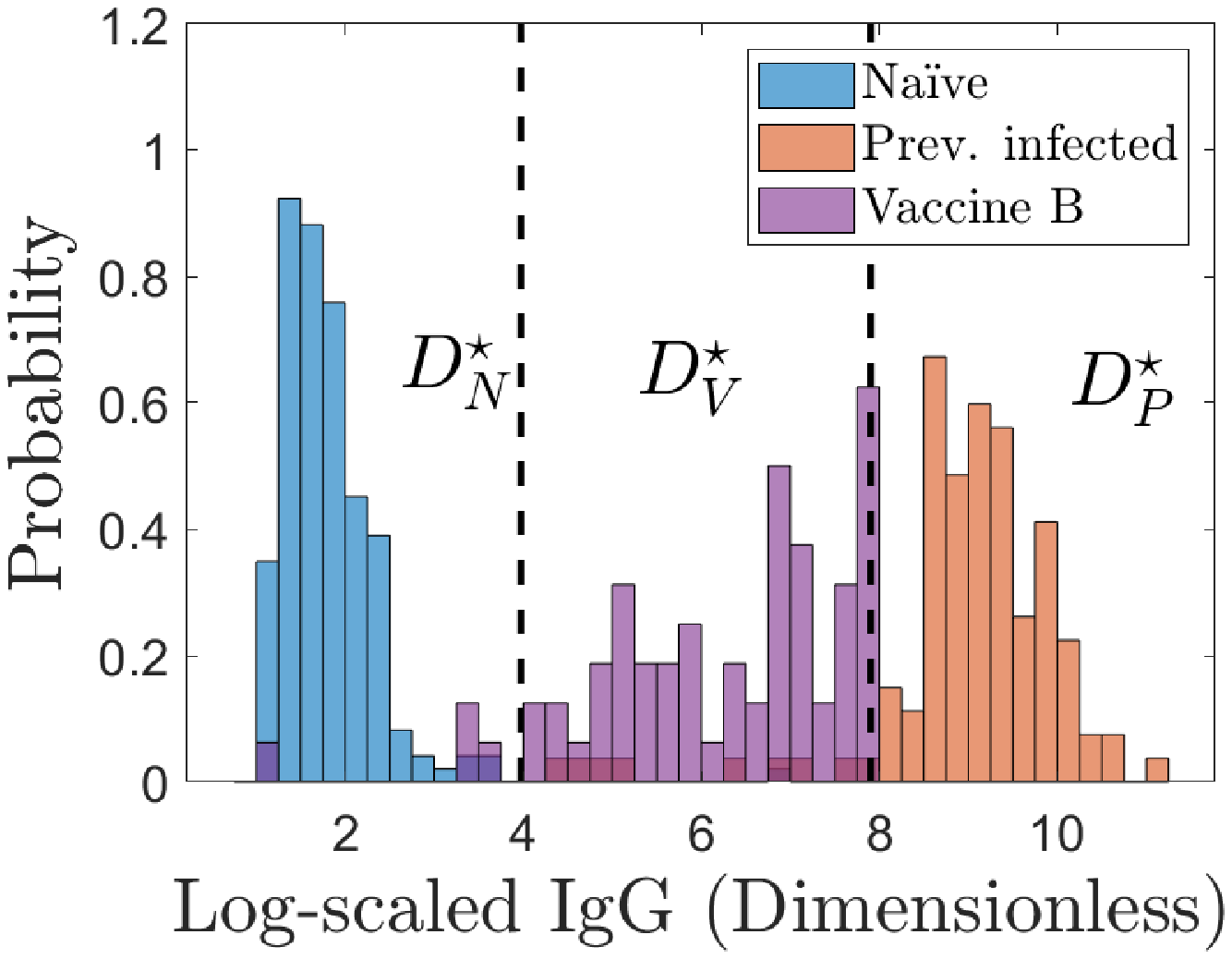}}
\caption{Training (a) and test (b) data with optimal decision thresholds using known (a) and estimated (b) prevalences for a Vaccine B vaccinated class. 
Vertical dashed lines indicate optimal decision boundaries. 
}
\label{fig:train_sup}
\end{figure}

\begin{figure}[H]
\centering
\subfloat[][Training data]{\includegraphics[scale=.5]{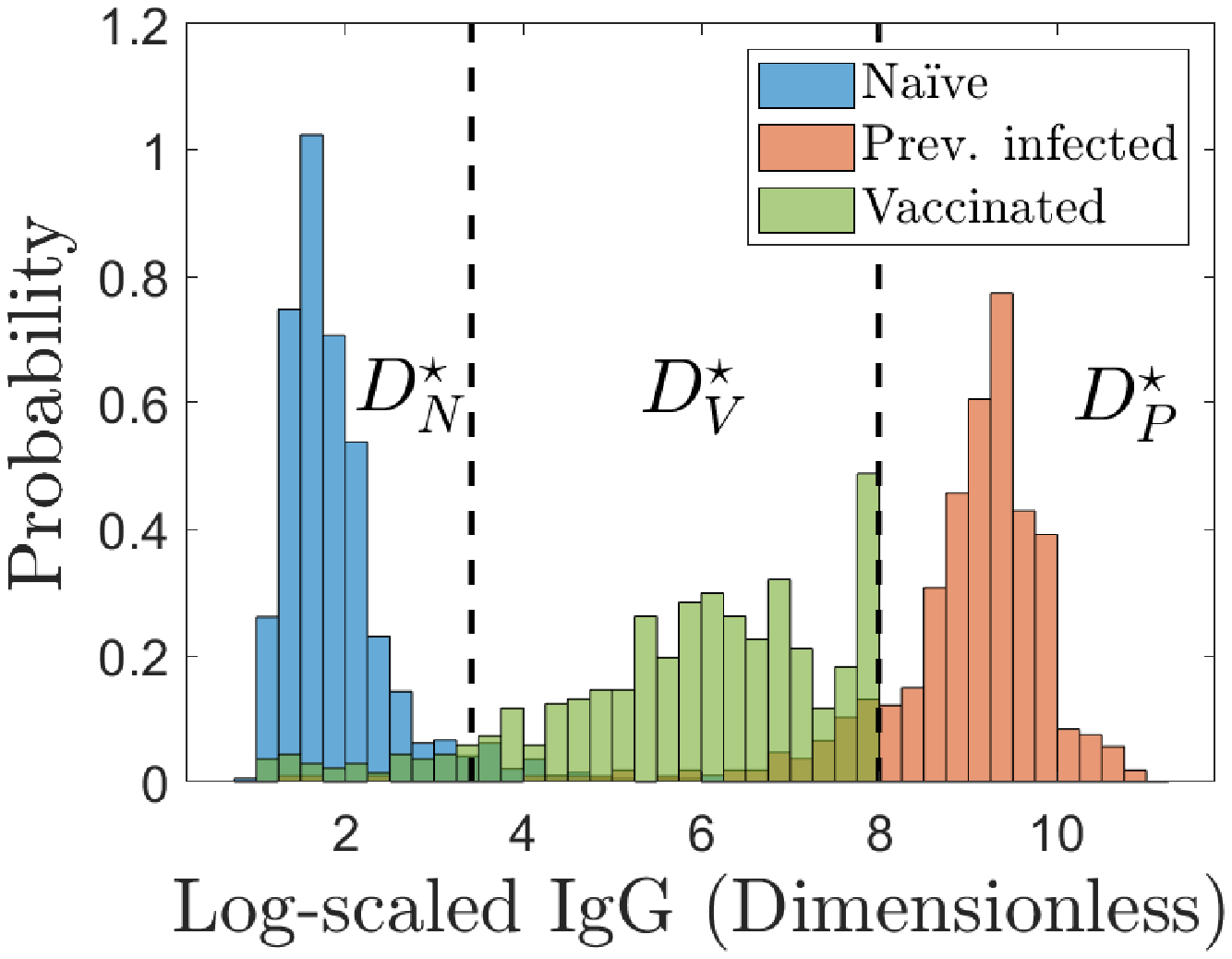}} 
\subfloat[][Test data]{\includegraphics[scale=.5]{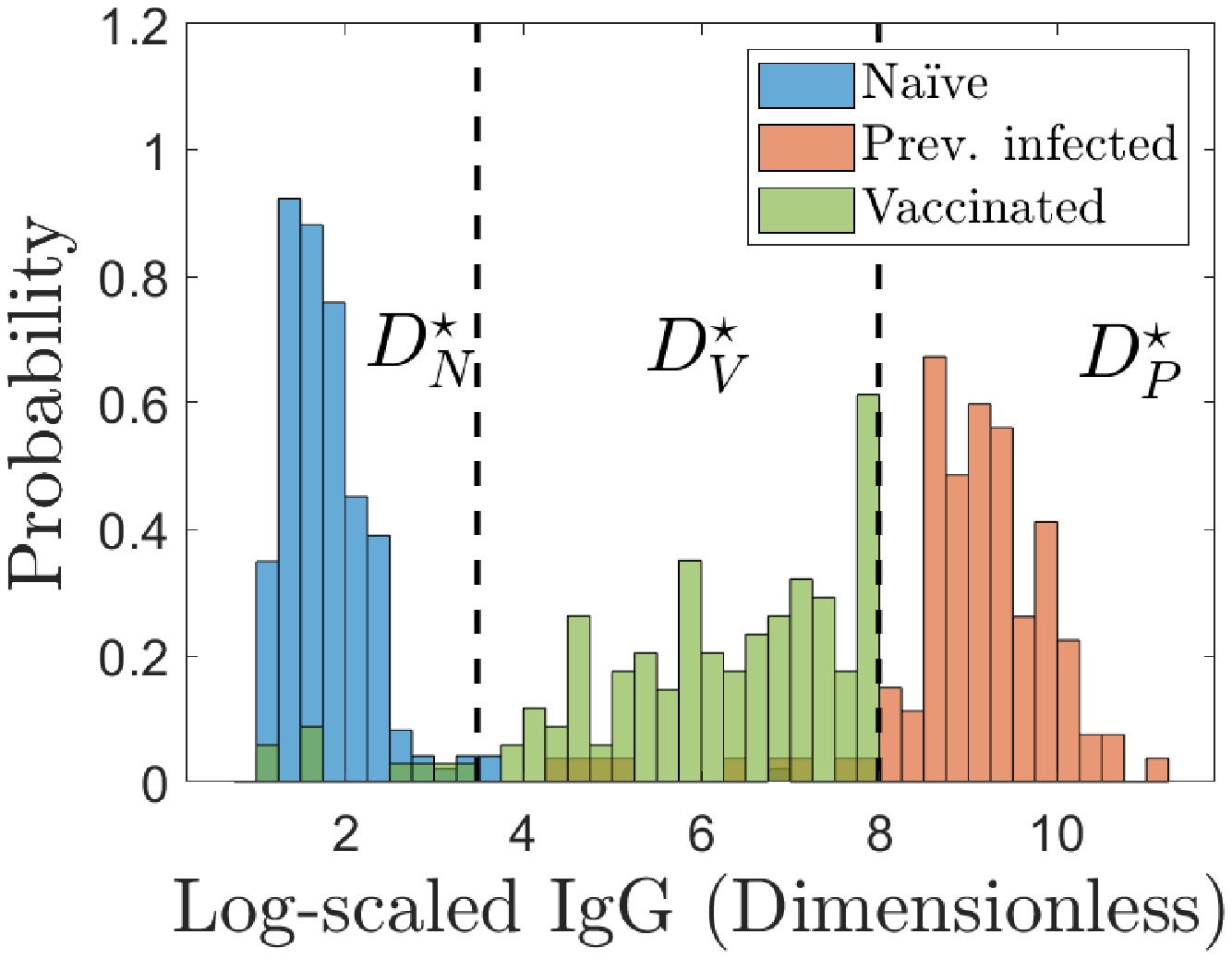}} 
\caption{Training (a) and test (b) data for a combined vaccinated class with optimal decision thresholds.  The training data is classified with known generalized prevalences and the test data using estimated  generalized prevalences. 
Vertical dashed lines indicate the optimal decision boundaries.
}
\label{fig:test_opt_sup}
\end{figure}

The optimal domains for the training and test data with a Vaccine B vaccinated class are shown in Figure \ref{fig:train_sup} and labeled as $D_N^{\star}$, $D_V^{\star}$, and $D_P^{\star}$; the same for a combined vaccinated class are shown in
Figure \ref{fig:test_opt_sup}. The training data is classified with known generalized prevalences and the test data using estimated  generalized prevalences. The optimal na{\"i}ve, 

\noindent  vaccinated, and previously infected domains are labeled $D_N^{\star}$, $D_V^{\star}$, and $D_P^{\star}$.   The optimal  decision boundary separating the vaccinated and previously infected class for both the Vaccine B and combined vaccinated samples lies at the upper truncation limit of the vaccinated data. Thus, there are no vaccinated samples misclassified as previously infected for these two cases. This may be caused by the accumulation of vaccinated samples in the largest bin; in contrast, the Vaccine A vaccinated class (Figure \ref{fig:train_test}) does not exhibit such a large spike. Our data-censored vaccinated PDFs account for this, and thus for large spikes, all samples at that truncation limit are sorted into the vaccinated class. 

\end{document}